\documentclass[prc,preprint,showpacs,floatfix,nofootinbib,tightenlines]{revtex4}
\usepackage{graphicx}  %
\usepackage{bm}  %

\usepackage{amssymb}

\begin{document}


\title{Neutron Densities from a Global Analysis of Medium Energy Proton
Nucleus Elastic Scattering}

\author{ B. C. Clark and L. J. Kerr}
\affiliation{Department of Physics \\
         The Ohio State University,\ \ Columbus, OH\ \ 43210}

\author{S. Hama}
\affiliation{Hiroshima University of Economics\\
 Hiroshima 731-0192, Japan}

%

%
\begin{abstract}
A new method for extracting neutron densities from intermediate
energy elastic proton-nucleus scattering observables uses
a global Dirac phenomenological (DP) approach
based on the Relativistic Impulse Approximation (RIA).
Data sets for $^{40}$Ca, $^{48}$Ca and $^{208}$Pb
in the  energy range from 500 MeV to 1040 MeV are considered.
The global fits are successful in reproducing the data
and in predicting data sets not included in the analysis.
Using this global approach, energy independent
neutron densities are obtained.
The vector point proton density distribution, $\rho^{p}_{v}$,
is determined from the empirical charge density
after unfolding the proton form factor. The other densities,
$\rho^{n}_{v}$, $\rho^{p}_{s}$, $\rho^{n}_{s}$,
are parametrized. 

This work provides energy independent values for the RMS neutron radius,
$R_n$ and the neutron skin thickness, $S_n$, in contrast to the
energy dependent values obtained by previous studies.
In addition, the results presented in paper show that the expected rms 
neutron radius and skin thickness for $^{40}$Ca is accurately reproduced.
The values of $R_n$ and $S_n$ obtained from the global fits
that we consider to be the most reliable are given as follows:
for $^{40}$Ca,  $3.314> R_n >3.310$ fm and
 $-0.063> S_n >-0.067$ fm; for $^{48}$Ca,
$3.459> R_n >3.413$ fm and $0.102> S_n >0.056$ fm; and
for $^{208}$Pb $5.550> R_n >5.522$ fm and $0.111> S_n >0.083$ fm.
These values are 
in reasonable agreement with nonrelativistic Skyrme Hartree-Fock models
and with relativistic Hartree-Bogoliubov models with density-dependent
meson-nucleon couplings. The results from the global fits 
for $^{48}$Ca and $^{208}$Pb are 
generally not in agreement with the usual relativistic mean-field 
models. 
%
%

\end{abstract}

\maketitle

\section{Introduction}
\label{sec:intro}

Determination of the proton and neutron densities, 
their root-mean-square radii, 
$R_p$ and $R_n$, and the neutron skin thickness, $S_n$ = $R_n$ - $R_p$, 
are critical to understanding many of the bulk properties of
matter \cite{Brown,Ring,Pom, Ring2}.  
Horowitz {\it et al.} 
have pointed out that there are substantial disagreement between 
theoretical values of $S_n$ \cite{horo1}. 
Furnstahl's recent analysis of neutron radii in the framework 
of mean-field models
shows that relativistic mean-field models
overestimate the values of $S_n$ \cite{Furn}. Parity violation 
electron scattering may  provide the experimental data to
resolve the  differences in the theoretical values \cite{jlab}, but 
is also useful to have alternative methods of obtaining neutron densities.
Reliable neutron densities  
are needed for atomic parity violation 
experiments \cite{horo1,For,Pol,Chen,Pol2},
the analysis of antiprotonic atoms \cite{trz}, in understanding 
the surface crust of neutron stars \cite{horo2}, and in 
extrapolation to proton-rich or neutron-rich nuclei that is important in
nuclear astrophysics \cite{WP2000}.  
In this work, we revisit the analysis of medium energy 
proton-nucleus elastic scattering data with the goal of obtaining 
reliable, energy independent neutron densities and the 
values of $R_n$ and $S_n$. The analysis of
elastic electron scattering which has resulted in reliable 
ground state charge densities has been a guiding light
for our work \cite{Frois}.

For a number of years we have used the 
relativistic impulse approximation (RIA) in the analysis of 
proton-nucleus elastic and inelastic scattering
\cite{lanny,newcc,sugie89,sugie90,lisa94,sugie99,sugie00,clark2} and 
the RIA-KDP(Kemmer-Duffin-Petiau) \cite{kemm,duff,pet}
for meson-nucleus elastic scattering \cite{prl,ku94,george,clark3,ku00}. 
These approaches produce relativistic optical potentials which result in
good agreement with medium-energy scattering observables.
The input to these calculations are the 
relativistic densities from Quantum Hadrodynamics (QHD)
\cite{serot86,serot92} and the elementary NN amplitudes from Arndt {\it et al.}
\cite{arndt}.  In recent work we use the modern EFT
densities \cite{clark2,clark3,FURNSTAHL97,RUSNAK97,RUSNAK97b}.

The seminal analysis of proton-nucleus elastic scattering data done by 
Ray and Hoffmann used both the RIA and the non-relativistic
KMT approach in their fits to get the observables from 300 MeV to 1040 MeV
\cite{RH}. 
Unfortunately neither approach produced energy-independent neutron densities.
In addition, some of the values of $S_n$ for $^{48}$Ca and $^{208}$Pb were 
negative, in contradiction to all nuclear structure calculations.  
Shlomo and Schaffer used the results from an analyses of
1 GeV proton elastic scattering from $^{40}$Ca and $^{48}$Ca
to obtain the skin thickness for $^{40}$Ca, $^{42}$Ca, $^{44}$Ca
and  $^{48}$Ca, see Table~2 in \cite{ss}.
Starodubsky and Hintz extracted the neutron densities from
elastic proton scattering of $^{206,207,208}$Pb at 650 MeV and
obtained $S_n$ of $(0.20 \pm 0.04)$ fm for $^{208}$Pb \cite{hintz}.
However, the energy-independence of the neutron densities in 
the work of Ref.~\cite{ss} or Ref.~\cite{hintz} was not addressed.
Recently Karataglidis {\it et al.}
have calculated proton and neutron elastic scattering 
from $^{208}$Pb and $^{40}$Ca targets 
at three energies 40 MeV, 65 MeV and 200 MeV. They used 
a model based on coordinate space nonlocal optical potentials 
using a full folding NN interactions with various Skyrme model 
ground-state densities \cite{Brown2}. 
For $^{208}$Pb they found that the SKM* model gave the
best agreement with proton and neutron elastic scattering
data at 40 MeV, 65 MeV and 200 MeV. Based on this the
authors suggest that  $S_n$ for $^{40}$Ca is $\sim$ -0.05 fm and 
$^{208}$Pb is $\sim$ 0.17 fm \cite{kara}.

In this paper, a new analysis of proton-nucleus elastic scattering is
used to obtain the neutron density. This work is
motivated by our considerable experience in obtaining high quality global 
proton-nucleus optical potentials from 20 MeV to 1040 MeV \cite{glob1}. 
The new method meshes the global approach with the RIA and proves to be  
successful in obtaining energy independent neutron densities.
The starting point is the RIA 
in its simplest form, which for spin-zero nuclei includes only scalar, 
vector and tensor terms. The tensor term is very small and is excluded;
as was done in the RIA analysis done by Ray and Hoffmann \cite{RH}.

We find that substantial progress in extracting the neutron densities
from proton-nucleus elastic scattering is made by using a global approach
focusing on the energy region where the RIA is capable of reproducing
experiment very well. We have obtained values of $S_n$ 
for $^{40}$Ca, $^{48}$Ca, and $^{208}$Pb which agree with 
nonrelativistic Skyrme models \cite{Brown,Brown2,Brown3} and 
relativistic Hartree-Bogoliubov model extended to include density 
dependent meson-nucleon couplings \cite{Ring2}. 
Our results for $^{48}$Ca and $^{208}$Pb 
are generally not in agreement with relativistic mean-field model, see 
Ref.~[6] and references therein.

The organization of this paper is as follows. Section~\ref{sec:glob}
describes the global method to obtain the neutron density.
Section~\ref{sec:results} discusses the results and the sensitivity
of the extracted neutron density, $R_n$ and $S_n$ to the input used in the
fitting procedure. The values of  $S_n$ and the neutron RMS
radii  $R_n$ for $^{40}$Ca, $^{48}$Ca, and $^{208}$Pb
for the various tests of the input to the model are given in this section.
The summary and conclusion is 
in Sect.~\ref{sec:sumcon}.

\section{RIA Global method for extracting the neutron density }
\label{sec:glob}

The RIA nuclear reaction formalism is used as the basis of 
global fits to medium-energy proton nucleus elastic scattering data. 
The input to the RIA consists of the Arndt NN amplitudes \cite{arndt}
and the point proton density, which is fixed from the 
charge distribution obtained 
from electron nucleus scattering. The neutron vector density,  
and the scalar proton and neutron densities are parametrized, 
resulting in good 
fits of p + A elastic scattering data between 500 MeV and 1040 MeV. 
Using the RIA as the basis for the global fits is a new approach 
and our results shows that it is a valid method for 
extracting neutron densities, $R_n$ and $S_n$.

In the global approach used in this work the form of the RIA vector and 
scalar and tensor optical potentials are given by

\begin{eqnarray}
U_s(r)& = &-\frac{P_{lab}}{(2\pi)^2 m}\sum_{j=p,n}\,\int_0^{\bar{q}}\,
4\pi q^2 dq\,\frac{R(q)}{R(0)}\,j_0(qr)\,F_s^{j}(q)\,
\tilde \rho_s^{j}(q),\\
U_v(r)& = &-\frac{P_{lab}}{(2\pi)^2 m}\sum_{j=p,n}\,\int_0^{\bar{q}}\,
4\pi q^2 dq\,\frac{R(q)}{R(0)}\,j_0(qr)\,F_v^{j}(q)\,
\tilde \rho_v^{j}(q), \\
\nonumber
U_t(r)& = &-\frac{P_{lab}}{(2\pi)^2 m}\sum_{j=p,n}\biggl[r\int_0^{\bar{q}}\,
4\pi q^2 dq \,\frac{R(q)}{R(0)}\,j_0(qr)\,F_t^{j}(q)\,
\tilde \rho_t^{j}(q) \\
      & + & \int_0^{\bar{q}}\, 4\pi q^2 dq\, {d \over dq}
 \biggl(\frac{R(q)}{R(0)} F_t^{j}(q)\biggr)j_1\,(qr)\tilde \rho_t^{j}
(q)\biggr],
\end{eqnarray}
where the fourier transforms of the density form factors are
\begin{eqnarray}
\tilde \rho_s^{j}(q)& = &\int d^3 r'\, e^{i \vec{q} \cdot \vec{r}'}
\rho_s^{j}(r'), \\
\tilde \rho_v^{j}(q)& = &\int d^3 r'\, e^{i \vec{q} \cdot \vec{r}'}
\rho_v^{j}(r'), \\
\tilde \rho_t^{j}(q)& = &\int d^3 r'\, e^{i \vec{q} \cdot \vec{r}'}
\frac{1}{r'}\, \rho_t^{j}(r').
\end{eqnarray}
The subscripts s,v,t refer to Lorentz scalar, vector (time-like), and
tensor quantities.  The superscript n and p refer to neutrons and protons,
$F(q)$ are the invariant NN amplitudes, and $R(q)$ is the kinematical
factor required to obtain the invariant NN amplitude in
the Breit frame \cite{RH}.
The value of the upper limit on the momentum transfer, ${\bar q}$, is
determined by the available on--shell NN data.  
Many studies conducted have demonstrated that
higher order corrections to this first order RIA-Dirac
optical model approach are negligible at the energies being
studied in this paper. See Ray {\it et al.} for more details of the 
RIA-Dirac calculations used for elastic scattering observables \cite{lanny}.

Elastic p + A data between 500 MeV and 1040 MeV for $^{40}$Ca,  
$^{48}$Ca and $^{208}$Pb form the data set.
The quality of the fits are good and the predictions
of data not in the data set is used to verify the procedure.
However, unlike the usual RIA, where we have generally 
used scalar, vector and tensor densities from the results of
relativistic effective field theory (EFT) calculations, we 
use the RIA as a basis for extracting the vector neutron density
as well as the two scalar densities using a global fitting procedure.

Next we describe the treatment of the vector and scalar densities.
The point proton density is fixed by using the results from electron 
scattering as follows,
\begin{eqnarray}
\rho_v^p(r)  & = &  \frac{1}{(2 \pi)^{3}}
\int d^3q e^{-i\vec{q}\cdot\vec{r}} \tilde{\rho}^p_v(q)
\end{eqnarray}
where, 
\begin{eqnarray}
\tilde{\rho}^p_v(q) & = &
\frac{\int d^3q e^{i\vec{q}\cdot\vec{r}}{\rho}_c(r)}{G(q)}.
\end{eqnarray}

\noindent Here $\rho_c(r)$ is the experimental charge density, 
$G(q)$ is the
proton form factor. The point proton density, $\rho_v^p(r)$ 
is normalized to $Z$ and 
the neutron density $\rho_v^n(r)$ is normalized to $N$.

The next step is to consider the model that will be used. 
After several years of testing a number of different
forms used to parametrize the neutron vector and scalar densities, we 
have chosen to model the vector neutron density and the scalar proton and
neutron densities using the cosh form (COSH) used in many of our 
global fits \cite{glob1}.
We have found the cosh form parametrization produced more stable
results for $R-n$ and $S_n$ than the three parameter
fermi (3PF) or the sum of gaussian (SOG) parametrizations.
This stability of the cosh was understood from the results from many global
fits using the all three forms. In global fits
we used four or five different momentum transfer ranges.
For example, if five momentum transfer ranges are used the fits are done 
from
0.0 $\mbox{fm}^{-1}$ to 1.5 $\mbox{fm}^{-1}$ to 0.0 $\mbox{fm}^{-1}$
to 3.5 $\mbox{fm}^{-1}$ in steps of 0.5 $\mbox{fm}^{-1}$.
In future work we will consider other parametrizations. 
 
The form of the cosh model in this work the superscript V stands for 
volume and S for surface terms,
\begin{eqnarray}
f^{V}(r,R,a) &=& \frac{\{\cosh[R/a]-1\}}{\{\cosh[R/a]+\cosh[r/a]-2\}}\\[8pt]
f^{S}(r,R,a) &=& \frac{\{\cosh[R/a]-1\}\{\cosh[r/a]-1\}}{\{\cosh[R/a]+
\cosh[r/a]-2\}^2}.
\end{eqnarray}
The vector neutron density is  
\begin{eqnarray}
 \rho_v^n(r,R_b,a_b) &=& \rho^{B}(r,R_b,a_b)  - \rho_v^p(r).
\end{eqnarray}
The density $\rho^{B}(r,R_b,a_b)$ is given by
\begin{eqnarray}
\rho^{B}(r,R_b,a_b) &\propto& f^{V}(r,R_b,a_b) + \alpha f^{S}(r,R_b,a_b),
\end{eqnarray}
\noindent and $\rho^{B}(r,R_b,a_b)$ is normalized to $A$.
There are two geometric parameters, $R_b$ and $a_b$, and the parameter, 
$\alpha$. 

The scalar proton and neutron densities are,
%
\begin{eqnarray}
\rho^{p}_s (r,R_s^{p},a_s^{p}) &\propto& f^{V}(r,R_s^{p},a_s^{p}) + 
\beta f^{S}(r,R_s^{p},a_s^{p})  
\end{eqnarray}
and 
\begin{eqnarray}
\rho^{n}_s (r,R_s^{n},a_s^{n}) &\propto& f^{V}(r,R_s^{n},a_s^{n}) + 
\gamma f^{S}(r,R_s^{n},a_s^{n}).  
\end{eqnarray}

\noindent 
Each of these densities contains three parameters, 
two geometry parameters, $R_b^{p}$, $R_b^{n}$, $a_b^{p}$, $a_b^{n}$
and the parameters $\beta$ and $\gamma$.
The 10th parameter, $P_{10}$, searched is given 
by $\int d^3 r \rho_s^p $ = $P_{10}$ Z.
A similar 11th parameter given by 
$ \int d^3 r \rho_s^n $ = $P_{11}$ N could also have been searched. 
However, the searches were more stable if the ratio 
$\frac{\int d^3 r \rho_s^p / Z} {\int d^3 r \rho_s^n / N}$, {\it i.e.},
the ratio of $\frac{P_{10}} {P_{11}}$, was fixed by ratio of the 
volume integrals per particle of 
the scalar proton and neutron densities from any one of the EFT models
given by Rusnak {\it et al.} \cite{RUSNAK97,RUSNAK97b}. 
Three different EFT densities are used, two are point coupling models, 
VA3 and FZ4, and one is a meson model, MA4.
The parameter $P_{10}$ is not sensitive to the EFT model chosen and the
final difference and the scalar proton and neutron densities are
not sensitive to the EFT model chosen even though the values of $S_n$
for N $\neq$ Z in these models differ widely. For example, values of $S_n$ for
$^{208}$Pb  are 0.332 $\mbox{fm}$ for VA3, 0.259 $\mbox{fm}$ 
for MA4 and 0.160 $\mbox{fm}$ for FZ4. 

As mentioned above we found that the model used in this paper 
was the best of many different models we tried. We will 
investigate other models in future work using this global approach.  


\section{Fitting Procedure and Results}
\label{sec:results}
It has long been known that using a global approach has been very useful in
obtaining NN amplitudes and the NN phase shifts. This is one of the 
reasons for using a global approach when the data set used is large and
usually correlated. In this 
work we use elastic p + A data between 500 MeV and 1040 MeV for $^{40}$Ca,  
$^{48}$Ca and $^{208}$Pb.  
For $^{40}$Ca there are five energies; 
497.5 MeV \cite{ca500,ca500a}, 613 MeV \cite{ca613}, 650 MeV \cite{ca650},
797.5 MeV \cite{ca800,ca800a}, and 1040 MeV \cite{ca1040,ca1040a}.
For $^{208}$Pb
there are five energies; 497.5 MeV \cite{ca500,pb500a}, 613 MeV \cite{ca613},
650 MeV \cite{pb650,pb650a}, 797.5 MeV \cite{pb800,pb800a}, 
and 1040 MeV \cite{pb1040}. 
However for $^{48}$Ca only three energies; 497.5 MeV \cite{ca500}, 
797.5 MeV \cite{c8a800}, and 1044 MeV \cite{ca1040,ca1040a} are available. 
In order to make predictions of data not included in the data set we 
remove one energy. For $^{40}$Ca and $^{208}$Pb the 650 MeV data has been
excluded, and for $^{48}$Ca 1044 MeV data has been excluded. 
The quality of the global fits to the data are good and the predictions 
for data not in the global data sets verify the procedure.  

In previous Dirac phenomenological (DP) global fitting, the data 
sets were cut at 100 degrees in the center of mass or at momentum 
transfer q at 3.0 $\mbox{fm}^{-1}$, 
whichever came first. 
In this work we do global fits using a variety of momentum transfer values
as discussed below.  In addition, as is discussed later, we also test the
sensitivity of the input to the model. 

In the global fitting we have used five momentum transfer ranges from 
 0.0 $\mbox{fm}^{-1}$ to 1.5 $\mbox{fm}^{-1}$ to 0.0 $\mbox{fm}^{-1}$ 
to 3.5 $\mbox{fm}^{-1}$ in steps of 0.5 $\mbox{fm}^{-1}$. 
Or four momentum transfer ranges from  0.0 $\mbox{fm}^{-1}$ to
2.0 $\mbox{fm}^{-1}$ to 0.0 $\mbox{fm}^{-1}$ to 3.5 $\mbox{fm}^{-1}$ 
again in steps of 0.5 $\mbox{fm}^{-1}$.  
The values of $S_n$ and $R_n$ are expected to
change with the value of the momentum transfer 
range used as the data sets are changed.
We find that the values of $S_n$ and $R_n$ for a given 
momentum transfer range (we use five momentum transfer ranges or 
four momentum transfer ranges) are    
clustered in a reasonably well constrained set of values for $S_n$ and 
$R_n$, we call these stable results.
The momentum transfer range for 0.0 $\mbox{fm}^{-1}$ to 1.5 $\mbox{fm}^{-1}$ 
is almost always the outlier, which is understandable as the data set 
is small and does not have the diffraction structure needed when fitting 
proton scattering from nuclear targets.
Of course in any such global approach the data sets are usually correlated. 
This is certainly true as the data sets sets with different 
momentum transfer ranges do overlap. Thus
a statistical analysis to obtain a mean value and
a standard deviation can not generally be used. However, we can 
determine the range of the values of $S_n$ and $R_n$ for each fit from
every  momentum transfer range,
and obtain the values for $S_n$ and $R_n$ which could be shown as 
a scatter plot, as is usually done when data sets used are correlated.
We have obtained the range of values for $S_n$ and $R_n$ 
for every test considered, while these ranges can not be interpreted as
resulting from a statistical analysis we found that doing such an 
analysis gives a useful guide to present the results. 

There are a number of features in the fitting procedures which could 
produce changes in the neutron density, $R_n$ and $S_n$.
As mentioned above, two ranges of momentum transfer cuts are used. In 
the following sensitivity tests we use both ranges. Generally the smaller 
range produces the smallest range of values for $R_n$ and $S_n$
so in most of the figures we show the larger range values 
for $R_n$ and $S_n$ which gives the most conservative results.

We test the results obtained when
the following features of the model used in the fitting procedures are 
changed. First the three different EFT models used in fixing the 
10th parameter are considered \cite{RUSNAK97,RUSNAK97b}.   
To investigate the effect of the $G(q)$ used in obtaining the 
point proton density we use two different forms they are
identified as  $G_1(q)$ from \cite{hol}
and  $G_2(q)$ from \cite {wir}. 
The set chosen for the Arndt NN amplitudes is input to the fitting
procedure, and we use sets FA00 and SM86 to find the sensitivity 
of the fits to this input \cite{arndt}. 
The data sets included in the fits have been changed to see if the 
results change significantly.  
Finally, we use the three different charge density models used in
the analysis of electron-nucleus elastic scattering data;  
the three parameter fermi (3PF), the sum of gaussian (SOG) and 
the fourier-bessel (FB) \cite{devries}. 
Thus each target we have done global fits which have used effective
field theory densities MA4, FZ4 and VA3, and for each 
three charge distributions are used. In addition two different form 
factors $G_1(q)$ and $G_2(q)$  and two sets of Arndt amplitudes, set 
FA00 and set SM86. Figures~1-5 show the results of these tests of the 
global fitting procedure.


The results for the $R_n$ and $S_n$ are given 
in three tables, found in
the Appendix; Table I for $^{40}$Ca, Table~II for $^{48}$Ca 
and Table~III for $^{208}$Pb.  

There are three general cases we use in testing the global fits:
Case 1 uses Arndt amplitudes set FA00 and the $G_1(q)$ form factor;
Case 2 uses Arndt amplitudes set FA00 and the $G_2(q)$ form factor;
and Case 3 uses Arndt amplitudes set SM86 and the $G_1(q)$ form factor.
In every case, we obtain good global fits for MA4, FZ4 and VA3 for
every charge distribution used; SOG, 3PF, and FB. As discussed above
the momentum cuts 
ranges used are 0.0 $\mbox{fm}^{-1}$ to 1.5 $\mbox{fm}^{-1}$ to
0.0 $\mbox{fm}^{-1}$ to 3.5 $\mbox{fm}^{-1}$ in steps of 
0.5 $\mbox{fm}^{-1}$ or we remove the 0.0 $\mbox{fm}^{-1}$ to 
1.5 $\mbox{fm}^{-1}$ set. 
For each case the one with five momentum transfer ranges or four 
momentum transfer ranges 
we calculate the mean and standard deviation for $R_n$ and $S_n$ for 
every global fit, this gives us an  average over all EFT models for a 
given charge distribution. This is  
denoted by $AVE_{EFT}$ in the tables. This is not to be taken as
a statistical error, it is however a convenient way to show the range of
the values in a consistent way. We do this rather than remove every outlier.  
Then for each case we show the combined values of the range for 
$R_n$ and $S_n$ for that case,
this gives the average over all charge distributions as well as
over all EFT models. These results are denoted as $AVE_{CD}$ and $AVE_{EFT}$ 
in the tables. Finally, we calculate the ranges of $R_n$ and $S_n$
for all three cases combined for each target. 

The last calculation, which combines all of the values 
of $R_n$ and $S_n$  for every test made, gives us the most 
conservative results.  
From these final test cases, shown at the bottom of the tables in the 
appendix, we get the following values using the five momentum transfer ranges 
for $^{40}$Ca,  $3.350> R_n >3.300$  fm and $-0.008> S_n >-0.080$ fm;
for $^{48}$Ca, $3.505> R_n >3.421$ fm and $0.148> S_n >0.058$ fm;  
for $^{208}$Pb, $5.589> R_n >5.513$ fm and $0.156> S_n >0.076$ fm. 
As mentioned earlier if we use the four momentum transfer ranges  
results are generally more clustered and  we obtain: 
for $^{40}$Ca, $3.350> R_n >3.304$ fm and $-0.006> S_n >-0.080$ fm;
for $^{48}$Ca, $3.485> R_n >3.417$ fm and $0.129> S_n >0.053$ fm; and 
for $^{208}$Pb, $5.561> R_n >5.513$ fm and $0.130> S_n >0.074$ fm.   
The values for $R_n$ and $S_n$ for both of the final case results are
quite similar. 

However, the authors found that the  
global fits using the Arndt NN amplitude set FA00 and the
form factor $G_1(q)$  gave the most stable results {\it i.e.} the
values did not vary very much,
see Tables~I, II and III in the Appendix. 
This is especially true for using the SOG charge distribution.
In this case the values of $R_n$ and $S_n$ using the five momentum transfer
ranges are:
for $^{40}$Ca, $3.318> R_n >3.308$ fm and $-0.059> S_n >-0.069$ fm; 
for $^{48}$Ca, $3.498> R_n >3.412$ fm and $0.141> S_n >0.055$ fm; and 
for $^{208}$Pb, $5.602> R_n >5.512$ fm and  $0.164> S_n >0.074$ fm.  
The values for $R_n$ and $S_n$ using the four momentum transfer
ranges are:
for $^{40}$Ca, $3.314> R_n >3.310$ fm and $-0.063> S_n >-0.067$ fm; 
for $^{48}$Ca, $3.459> R_n >3.413$ fm and $0.102> S_n >0.056$ fm; and
for $^{208}$Pb, $5.550> R_n >5.522$ fm and $0.111> S_n >0.083$ fm.  

The results using the form factor $G_2(q)$ are quite similar as well as the
case using  form factor $G_1(q)$ but with the NN set SM86.
These values as well as the values using the obtained from the 
final case values are in agreement with nonrelativistic Skyrme models 
\cite{Brown,Brown2,Brown3}, the relativistic Hartree-Bogoliubov
model extended to include density dependent meson-nucleon couplings
\cite{Ring2} and the recent analysis of antiproton atoms \cite{trz}.
Our results are generally not in agreement with relativistic mean-field models.

\begin{figure}[p!]
\begin{center}
\includegraphics[width=4.6in,angle=0,clip=true]{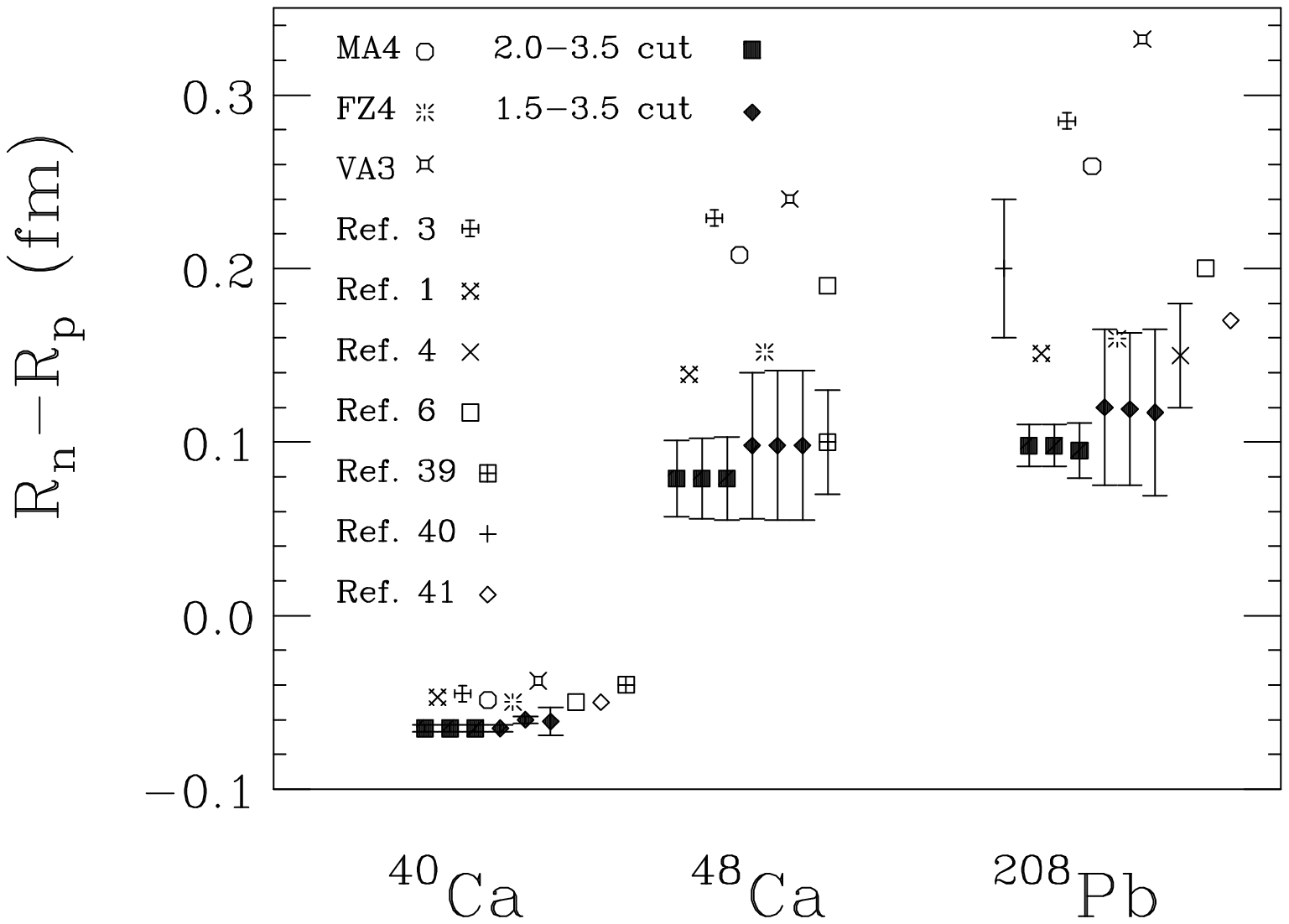}
\caption{The ranges, shown as bars, 
of the skin thickness for $^{40}$Ca, 
$^{48}$Ca and $^{208}$Pb
from the global fits using the SOG charge distribution, set FA00
Arndt NN amplitudes, proton form factor $G_1(q)$, and the three EFT models,
MA4, FZ4 and VA3 are shown by the filled boxes, for the 
four momentum transfer ranges and the filled diamonds for 
the five momentum transfer ranges.
Several theoretical $S_n$ values from 
Refs.~ \cite{Brown,Pom,Ring2,Furn,RUSNAK97,RUSNAK97b,ss,hintz,kara} are 
also shown by using various symbols.}
\label{fig:diffsog1}
\vspace*{.1in}
\includegraphics[width=4.6in,angle=0,clip=true]{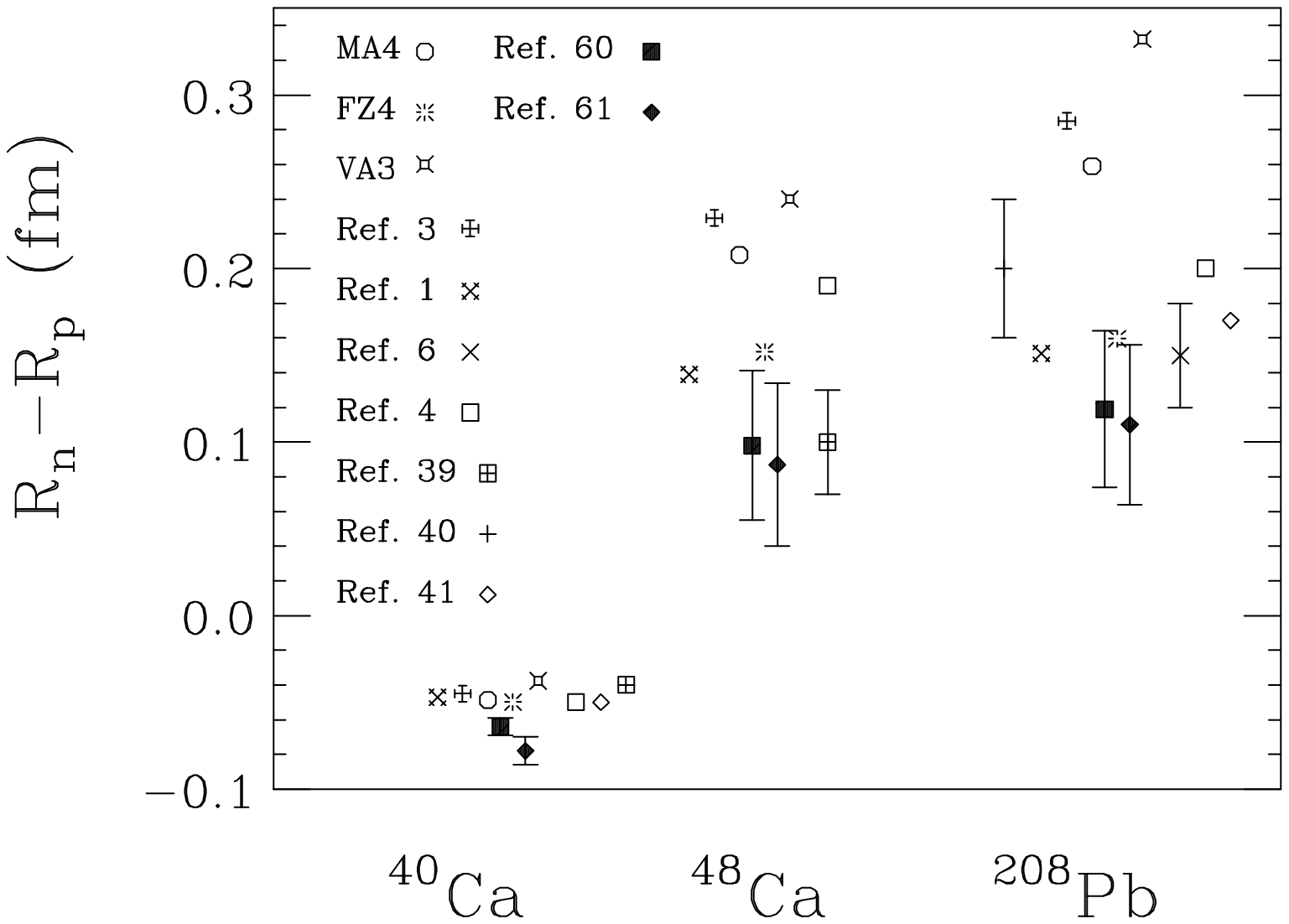}
\caption{The skin thickness 
with different proton form factors.
Results for $^{40}$Ca, $^{48}$Ca and $^{208}$Pb from
the global fitting procedure shows the $AVE_{EFT}$ models values for the 
SOG charge distribution, set FA00
Arndt NN amplitudes, the five momentum transfer ranges.
We compare $G_1(q)$ from Ref.~\cite{hol} shown as filled boxes and
$G_2(q)$ from Ref.~\cite{wir} shown as filled diamonds.
The same theoretical values for $S_n$ shown in Fig.~\ref{fig:diffsog1} are
also shown.}
\label{fig:diffdpoff1}
\end{center}
\end{figure}

Next we discuss the results of each test of the model separately.
Figure~\ref{fig:diffsog1} shows the effect of the two different 
momentum transfer ranges 
as well as the difference in the three EFT models used to fix parameter 10. 
The filled boxes (the smaller momentum transfer range) and diamonds (the
larger momentum transfer range) are the values from the three different
EFT models used. The charge density used in this 
case was the SOG for all targets, the proton form factor 
$G_1(q)$, and the set FA00 Arndt NN amplitudes.
This figure shows that the results from the two different 
momentum ranges overlap and that the EFT model used does not affect the results.
The figure also shows several 
theoretical results from 
Refs.~\cite{Brown,Pom,Ring2,Furn,RUSNAK97,RUSNAK97b,ss,hintz,kara}.
The comparison between theory and range of the values from the 
fits shows that all theoretical $S_n$ values agree with the results of 
the global fits for  $^{40}$Ca. However, the theoretical $S_n$ values for 
$^{48}$Ca and $^{208}$Pb are somewhat larger than the global fits. 


Next we consider the results when 
two different proton form factors, $G_1(q)$ from Ref.~\cite{hol}
and $G_2(q)$ from Ref.~\cite{wir}, are used in
obtaining the vector point proton density for a given charge 
distribution model. The five momentum transfer ranges,
the values for the $AVE_{EFT}$ models for the SOG model charge 
distribution and set FA00 Arndt NN amplitudes are used and shown in 
Figure~\ref{fig:diffdpoff1}. 

\begin{figure}[p!]
\begin{center}
\includegraphics[width=4.4in,angle=0,clip=true]{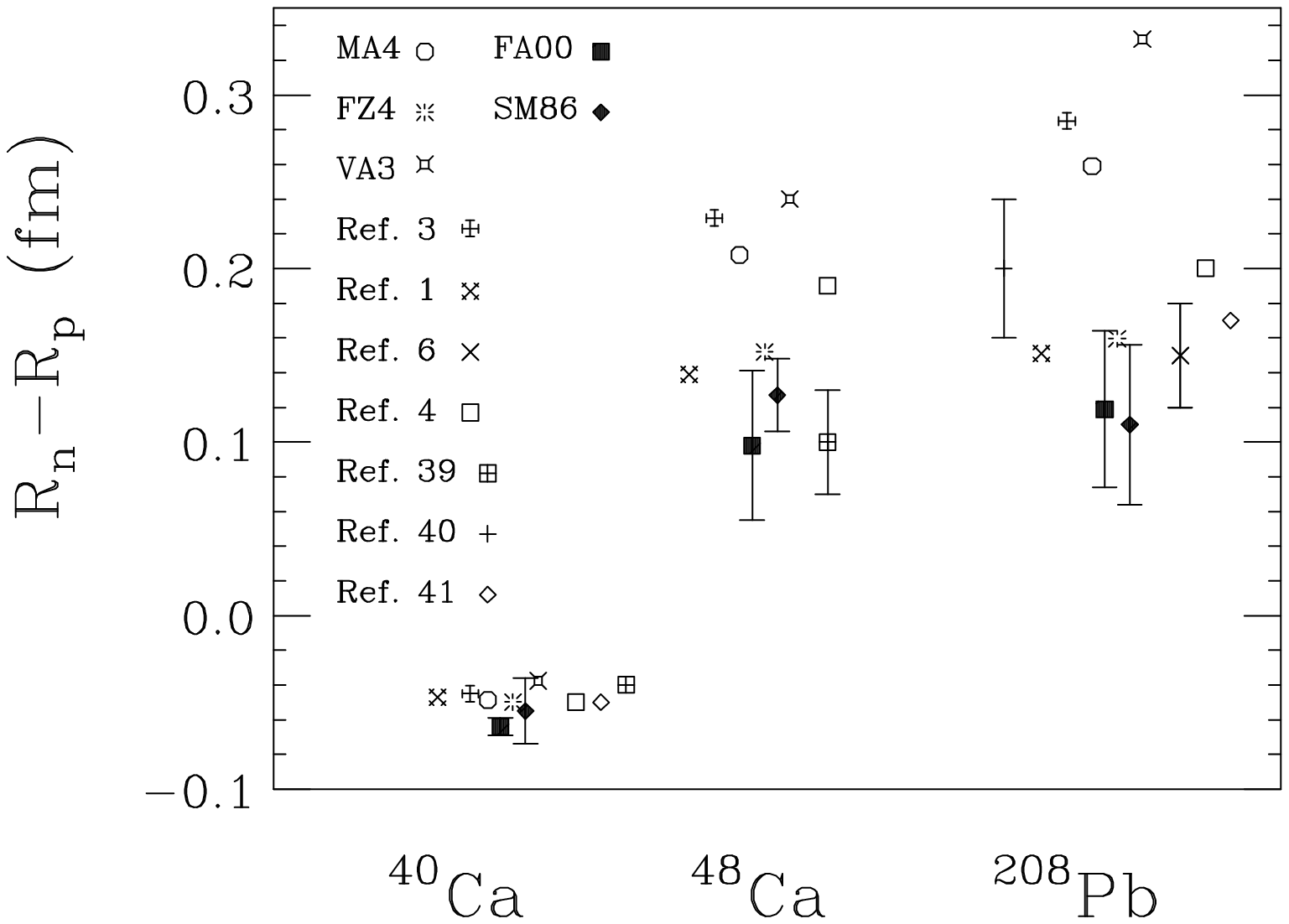}
\caption{The skin thickness for $^{40}$Ca, $^{48}$Ca and $^{208}$Pb
due to different sets of Arndt NN amplitudes. 
The global results shown are the $AVE_{EFT}$ models for 
SOG charge distribution, 
the five momentum transfer ranges,  and proton form factor $G_1(q)$.
Set FA00 is shown as filled boxes and set SM86 is shown as 
filled diamonds.
The same theoretical values for $S_n$ shown in Fig.~\ref{fig:diffsog1} are 
also shown.}
\label{fig:diffFA00SM86}
\vspace*{.1in}
\includegraphics[width=4.4in,angle=0,clip=true]{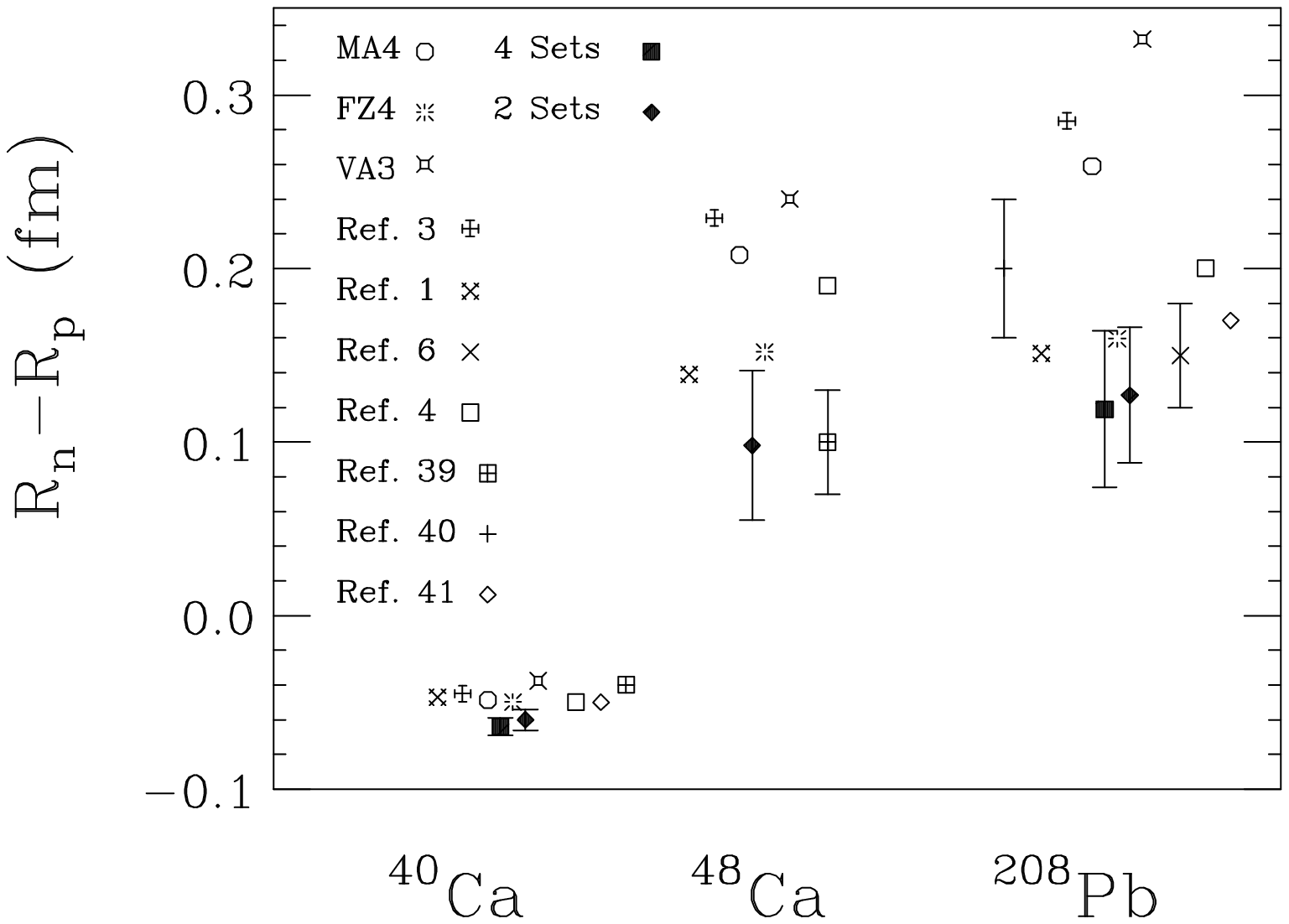}
\caption{The skin thickness for $^{40}$Ca, $^{48}$Ca 
and $^{208}$Pb when different data sets included in the fits, 
one has four energies and the other has two energies.
The results of the global procedure uses the $AVE_{EFT}$ models 
and the SOG charge distribution, the five 
momentum transfer ranges,
the proton form factor $G_1(q)$, set FA00 Arndt NN amplitudes. 
The results for four-energy sets are shown by filled boxes and
the results for two-energy data sets by filled diamonds. The data set for
$^{48}$Ca has only three energies so only has two data sets
are used in the fits. The same theoretical values for $S_n$ shown in
Fig. \ref{fig:diffsog1} are also shown.}
\label{fig:difftwosets}
\end{center}
\end{figure}

Investigating the sensitivity to the set of Arndt NN amplitudes 
used is done by comparing sets FA00 and SM86.
The larger momentum transfer range,
the averaged EFT models for the SOG model charge distribution, and the 
proton form factor $G_1(q)$ are used and shown in   
Figure~\ref{fig:diffFA00SM86}.

In order to check the sensitivity due to the data sets included in the
fit we have done fits using only two data sets (497.5 MeV and 797.5 MeV)
for $^{40}$Ca and $^{208}$Pb. 
The results agree very well with the same case using four data sets
as is shown in Fig.~\ref{fig:difftwosets}. 
The values of  $S_n$ for the two data set case are for $^{40}$Ca,
$-0.054> S_n >-0.066$ fm; and for $^{208}$Pb $0.166> S_n >0.088$ fm.
These results agree very well with the values for the four data set case
$^{40}$Ca, $-0.059> S_n >-0.069$ fm and $^{208}$Pb $0.164> S_n >0.074$ fm.
This encourages us to extend our global procedure to nuclei that
have at least two data sets in the medium energy range.

\begin{figure}[p!]
\begin{center}
\includegraphics[width=4.6in,angle=0,clip=true]{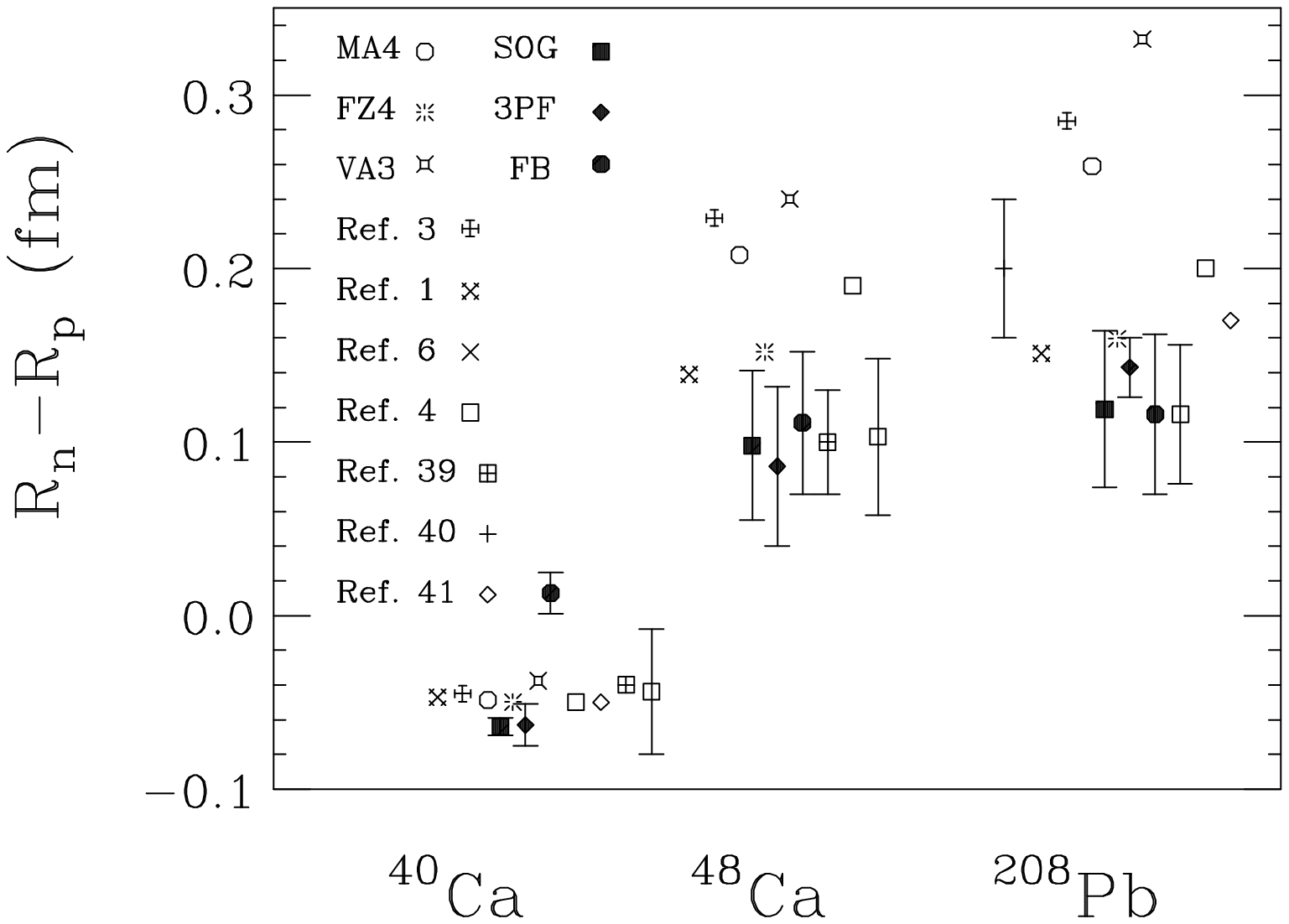}
\caption{The figures shows the change in the 
skin thickness for $^{40}$Ca, $^{48}$Ca and $^{208}$Pb
due to different charge distribution models.
The global fits use set FA00 Arndt NN amplitudes, 
the five momentum transfer ranges, and the
proton from factor $G_1(q)$. The results for the $AVE_{EFT}$ SOG 
are shown by filled boxes, 
the $AVE_{EFT}$ 3PF are shown by filled diamonds and the $AVE_{EFT}$ FB 
are shown by filled circles. The results for all effective field 
theory cases and for all three charge distributions {\it i.e.} the
results from the final cases are given by the open squares.
The same theoretical values for $S_n$ shown in 
Fig.\ref{fig:diffsog1} are also shown.}
\label{fig:diffcd1}
\end{center}
\end{figure}

The sensitivity of $S_n$ and $R_n$ to the three different charge 
distributions, three parameter fermi (3PF), 
sum of gaussian (SOG) and fourier-bessel (FB) obtained from
Ref.~\cite{devries} which are used in the fitting 
procedure is shown in Fig.~\ref{fig:diffcd1} and given in detail in 
Tables I, II and III. The five momentum transfer ranges,
the proton form factor $G_1(q)$ and the set FA00 Arndt NN  
amplitudes are used.
For $^{208}$Pb and $^{48}$Ca there is little 
difference in the values of $S_n$ and $R_n$ 
for the different charge densities. We note that the
RMS radii for $^{208}$Pb and $^{48}$Ca for
these three charge density are about the same \cite{devries}. 
Two of the three $^{40}$Ca charge distributions also have almost 
identical charge RMS radii, 3PF 3.482(25)fm , SOG and 3.479(3)fm but the  
RMS radius for the FB, 3.450(10)fm, is considerably smaller (see
Ref.~\cite{devries}). 
The result is that for $^{40}$Ca the difference in the value of
$S_n$ is pronounced. As shown in Fig.~\ref{fig:diffcd1} the 
FB charge distribution 
has range of  $S_n$ values that goes from  positive to 
negative but the range of the $S_n$ values 3PF and SOG  are all negative.  
We attribute this to the smaller RMS radius for the FB 
charge distribution.
The values of $R_p$, $R_n$ and $S_n$ should not depend heavily on
the momentum transfer ranges, the set of amplitudes, the form factors,
and the charge distribution. In fact, the only case that 
does not overlap is the FB case for $^{40}$Ca. 
The results for all three of the charge distributions are
given in Tables~I-III in the Appendix.    

In this global analysis we obtain many densities from the various
cases used in checking the results of the fits.
Figures~6 and 7 show the vector point proton density distribution, 
$\rho^{p}_{v}$ and $r^{2} \rho^{p}_{v}$ and 
the vector point neutron density distribution, 
$\rho^{n}_{v}$ and $r^{2} \rho^{n}_{v}$  for 
$^{40}$Ca, $^{48}$Ca, and $^{208}$Pb. 
The SOG charge distribution and the 
$G_1(q)$ form factor were used to obtain the vector point proton density.
The neutron density also depends on the charge distribution chosen and
the form factor used to obtain the vector point proton density as well as 
the other input to a given global fit, the NN amplitudes, 
and the EFT density used to fix the ratio of $\frac{P_{10}} {P_{11}}$. 
In  Figs.~6 and 7 the Arndt NN amplitude set FA00,  
the SOG charge distribution, the momentum transfer range was
 0.0 $\mbox{fm}^{-1}$ to  3.0 $\mbox{fm}^{-1}$, proton from factor $G_1(q)$,
and all  three EFT cases, MA4, FZ4 and VA3 are used. It is clear that the
neutron densities overlap showing that the use of different EFT cases
does not have any significant impact on the global fit. Tables of the
densities and the parameters are available from the authors \cite{author}.

\begin{figure}[p!]
\begin{center}
\includegraphics[width=5.0in,angle=180,clip=true]{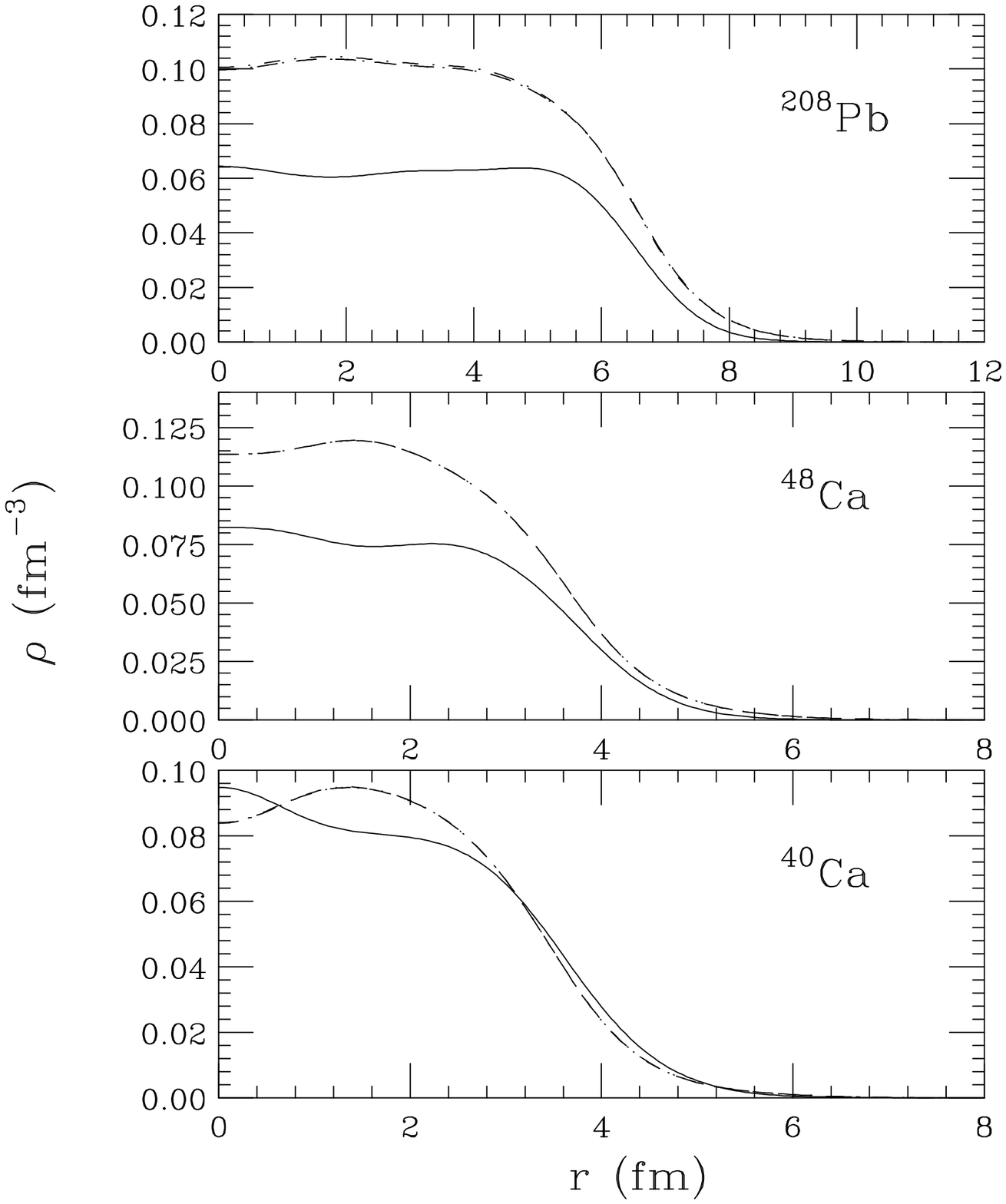}
\caption{The vector point proton density distribution shown by the solid 
curve. The three neutron density distribution densities using the  
EFT densities, MA4, FZ4 and VA3 
are shown by the dashed line for case MA4, the dots for case FZ4,
and the dotdashed for case VA3.  All  density  are 
obtained from the global fit as discussed in the manuscript.}
\label{fig:den}
\end{center}
\end{figure}

\begin{figure}[p!]
\begin{center}
\includegraphics[width=5.0in,angle=180,clip=true]{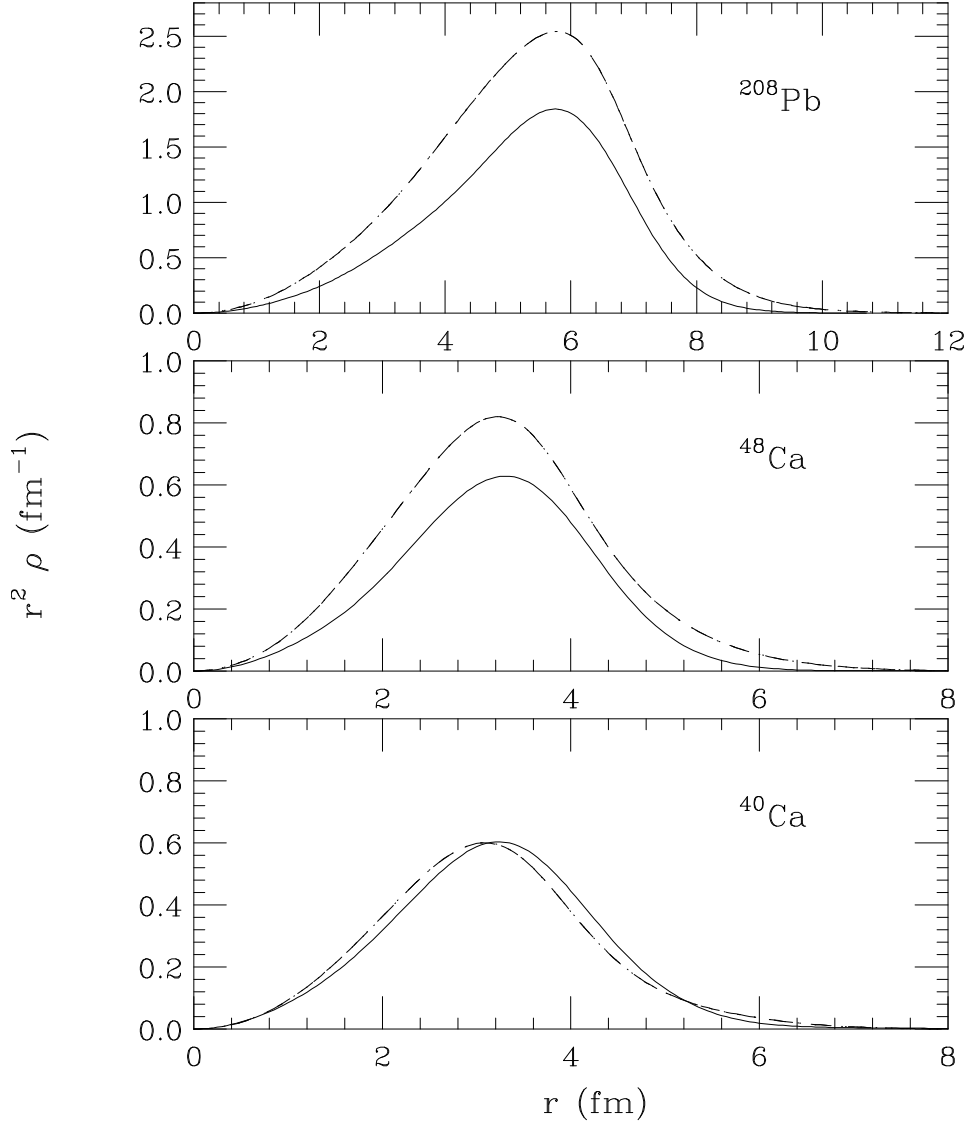}
\caption{The vector point proton density distribution shown by the solid 
curve. The three neutron density distribution densities using the  
EFT densities, MA4, FZ4 and VA3 
are shown by the dashed line for case MA4, the dots for case FZ4,
and the dotdashed for case VA3.  All  density  are 
multiplied by the radius squared and obtained from the global fit as 
discussed in the manuscript.}
\label{fig:den2}
\end{center}
\end{figure}

This same input as in Figs~6-7 is also
used in the Figs.~8-9 which shows the vector proton point density and
the scalar proton $\rho^{p}_{s}$ and $r^{2} \rho^{p}_{s}$, 
and in Figs.~10-11 for the vector point density and
the scalar neutron $\rho^{n}_{s}$ and $r^{2} \rho^{n}_{s}$.
Figures~8-11 show that the phenomenology gives sensible
scalar densities that are not unphysical.

\begin{figure}[p!]
\begin{center}
\includegraphics[width=5.0in,angle=180,clip=true]{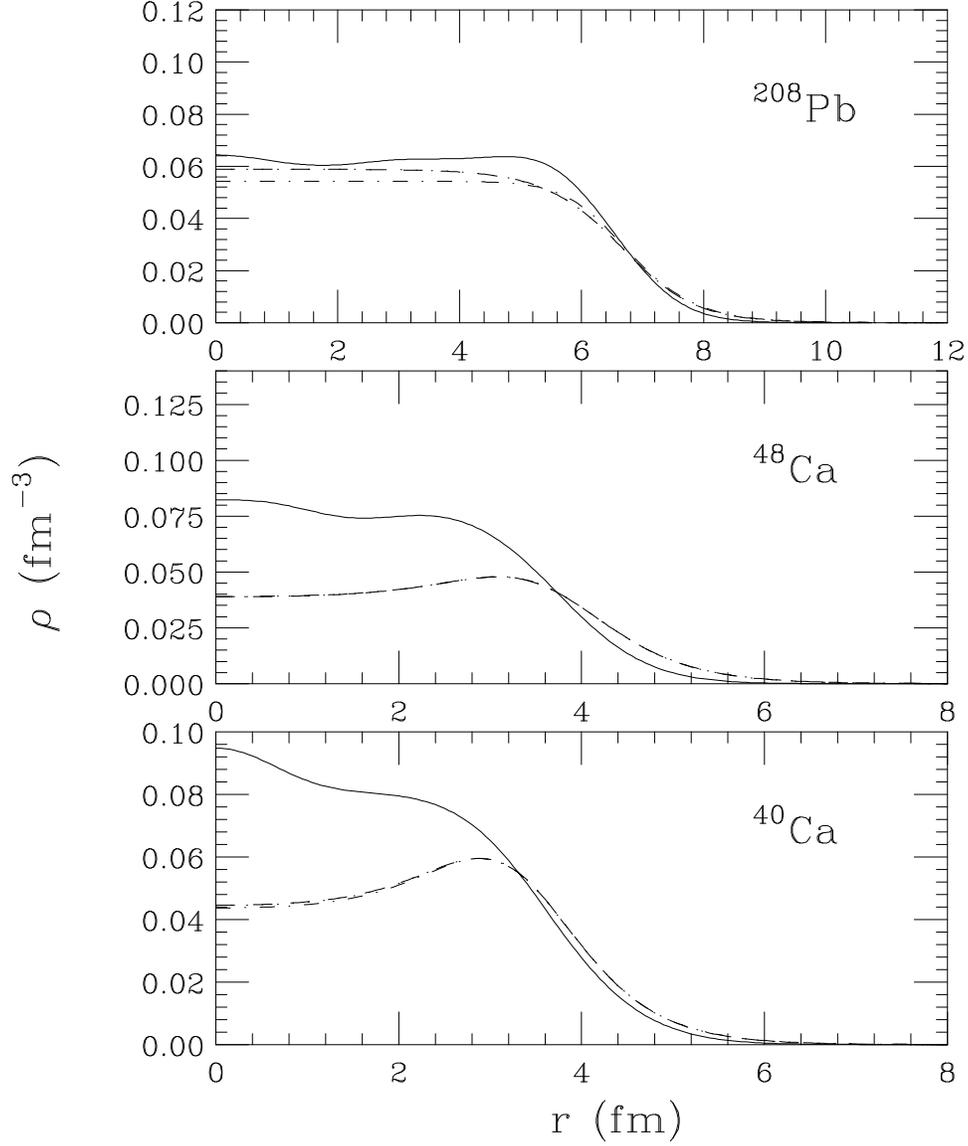}
\caption{The vector point proton density distribution shown by the solid 
curve. The three scalar proton density distribution densities using the  
EFT densities, MA4, FZ4 and VA3 
are shown by the dashed line for case MA4, the dots for case FZ4,
and the dotdashed for case VA3.  All  density  are 
obtained from the global fit as discussed in the manuscript.}
\label{fig:den3}
\end{center}
\end{figure}

\begin{figure}[p!]
\begin{center}
\includegraphics[width=5.0in,angle=180,clip=true]{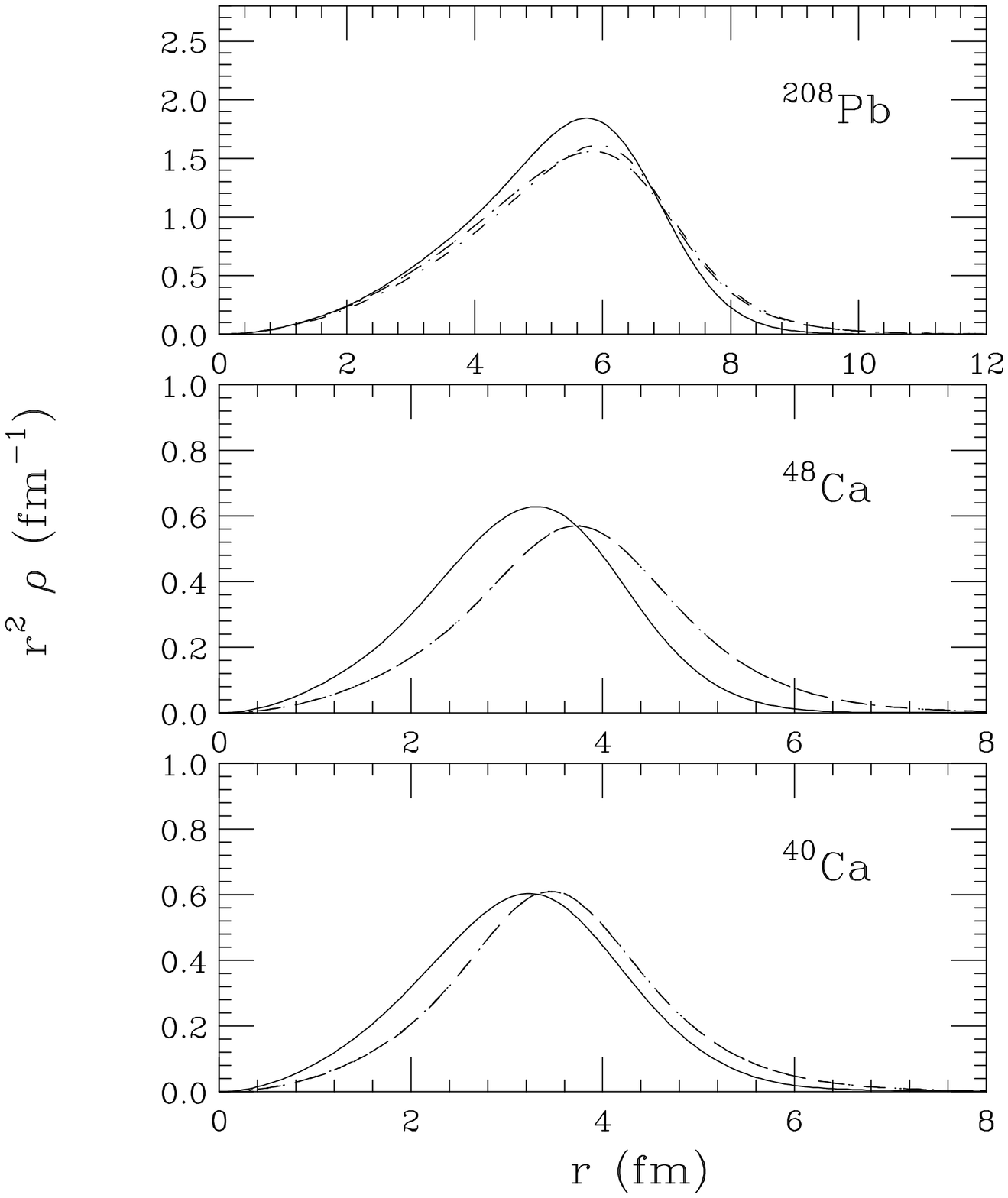}
\caption{The vector point proton density distribution shown by the solid 
curve. The three scalar proton density distribution densities using the  
EFT densities, MA4, FZ4 and VA3 
are shown by the dashed line for case MA4, the dots for case FZ4,
and the dotdashed for case VA3.  All  density  are 
multiplied by the radius squared and obtained from the global fit as 
discussed in the manuscript.}
\label{fig:den4}
\end{center}
\end{figure}

\begin{figure}[p!]
\begin{center}
\includegraphics[width=5.0in,angle=180,clip=true]{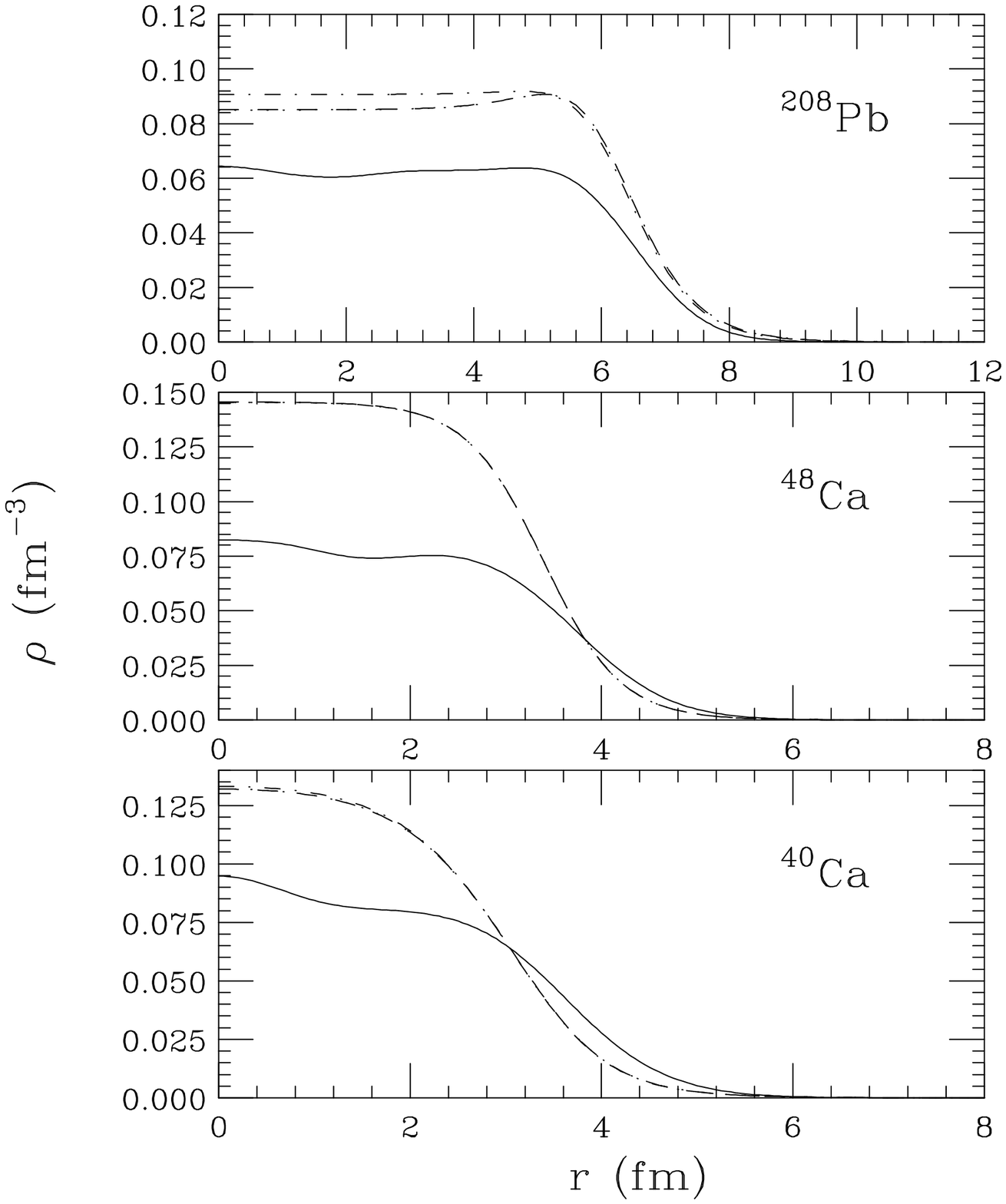}
\caption{The vector point proton density distribution shown by the solid 
curve. The three neutron scalar density distribution densities using the  
EFT densities, MA4, FZ4 and VA3 
are shown by the dashed line for case MA4, the dots for case FZ4,
and the dotdashed for case VA3.  All  density  are 
obtained from the global fit as discussed in the manuscript.}
\label{fig:den5}
\end{center}
\end{figure}

\begin{figure}[p!]
\begin{center}
\includegraphics[width=5.0in,angle=180,clip=true]{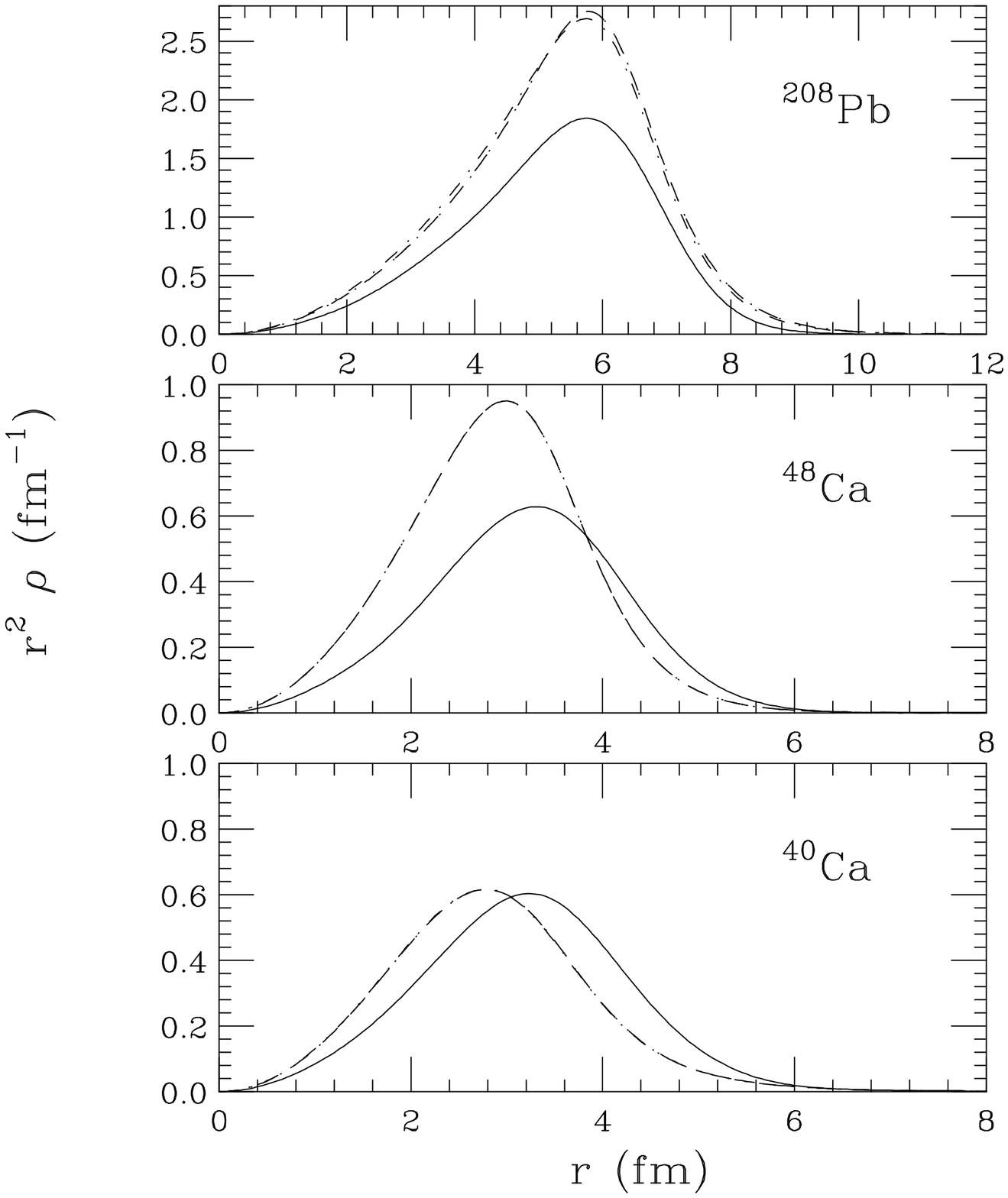}
\caption{The vector point proton density distribution shown by the solid 
curve. The three neutron scalar density distribution densities using the  
EFT densities, MA4, FZ4 and VA3 
are shown by the dashed line for case MA4, the dots for case FZ4,
and the dotdashed for case VA3.  All densities are 
multiplied by the radius squared and obtained from the global fit as 
discussed in the manuscript.}
\label{fig:den6}
\end{center}
\end{figure}

Figures~12 and 13 show the observables for one of the energies included in  
the global fit for $^{40}$Ca and a prediction
of an energy not included in the global fit.
Figure~12 shows the fit for $^{40}$Ca at 497.5 MeV and  
Fig.~13  shows the
prediction for the $^{40}$Ca 650 MeV spin observables; no cross section 
data is available.  Figures~14 and 15 show the observables for
one of the energies included in 
the global fit for $^{48}$Ca and the results from a prediction
of an energy not included in the global fit.
Figure~14 shows the fit for $^{48}$Ca at 497.5 MeV
and the prediction of Q at that energy. 
Figure~15 shows the prediction 
for $^{48}$Ca at 1044 MeV. Figures~16 and 17 show the observables for
one of the energies included in 
the global fit for $^{208}$Pb and the results from a prediction
of an energy not included in the global fit.
Figure~16 shows the fit for $^{208}$Pb at 497.5 MeV and 
Fig.~17 shows the prediction 
for the 650 MeV  observables for this target. The observables shown in
Figs.~12-17 all used the Arndt NN amplitude set FA00,  
the SOG charge distribution, the proton from factor $G_1(q)$,
the momentum transfer range from 
0.0 $\mbox{fm}^{-1}$ to  3.0 $\mbox{fm}^{-1}$,
and the EFT case MA4.  

While the figures of the observable are small we have magnified them
and we find that the
heights of the diffractive maxima and the angular positions of the
minima and maxima are very well reproduced at each energy with no
systematic energy dependent discrepancies.  These are the most
critical features of the data which determine rms radii.  In fact,
precision fits to $A_y$ and Q are not as important for
determining neutron radii but our fits are quite reasonable.
The magnified figures are available from the authors \cite{author}.

\begin{figure}[p!]
\begin{center}
\includegraphics[width=5.0in,angle=180,clip=true]{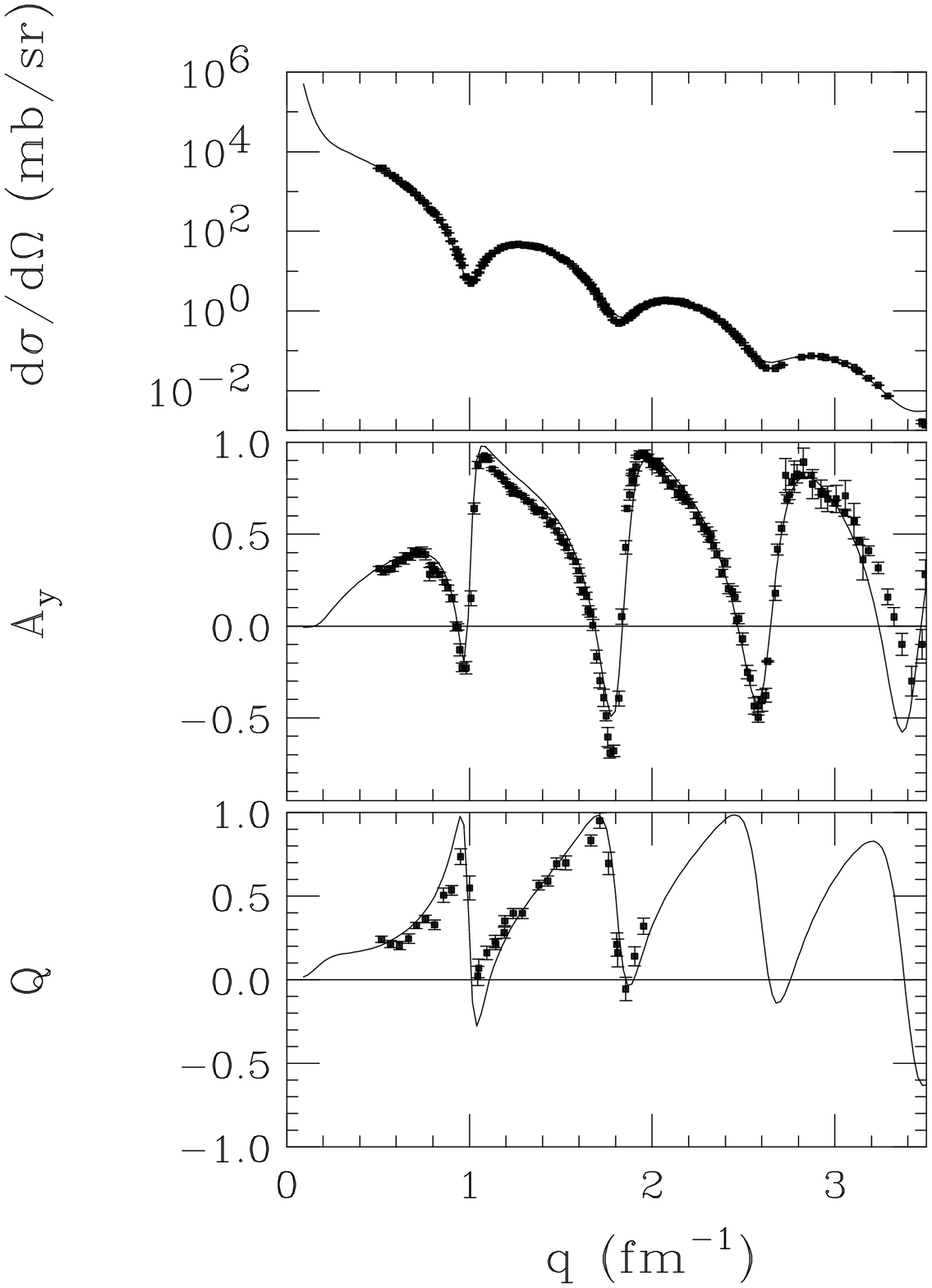}
\caption{Results of the global analysis for $^{40}$Ca at 497.5 MeV.}
\label{fig:ca497spin}
\end{center}
\end{figure}

\begin{figure}[p!]
\begin{center}
\includegraphics[width=5.0in,angle=180,clip=true]{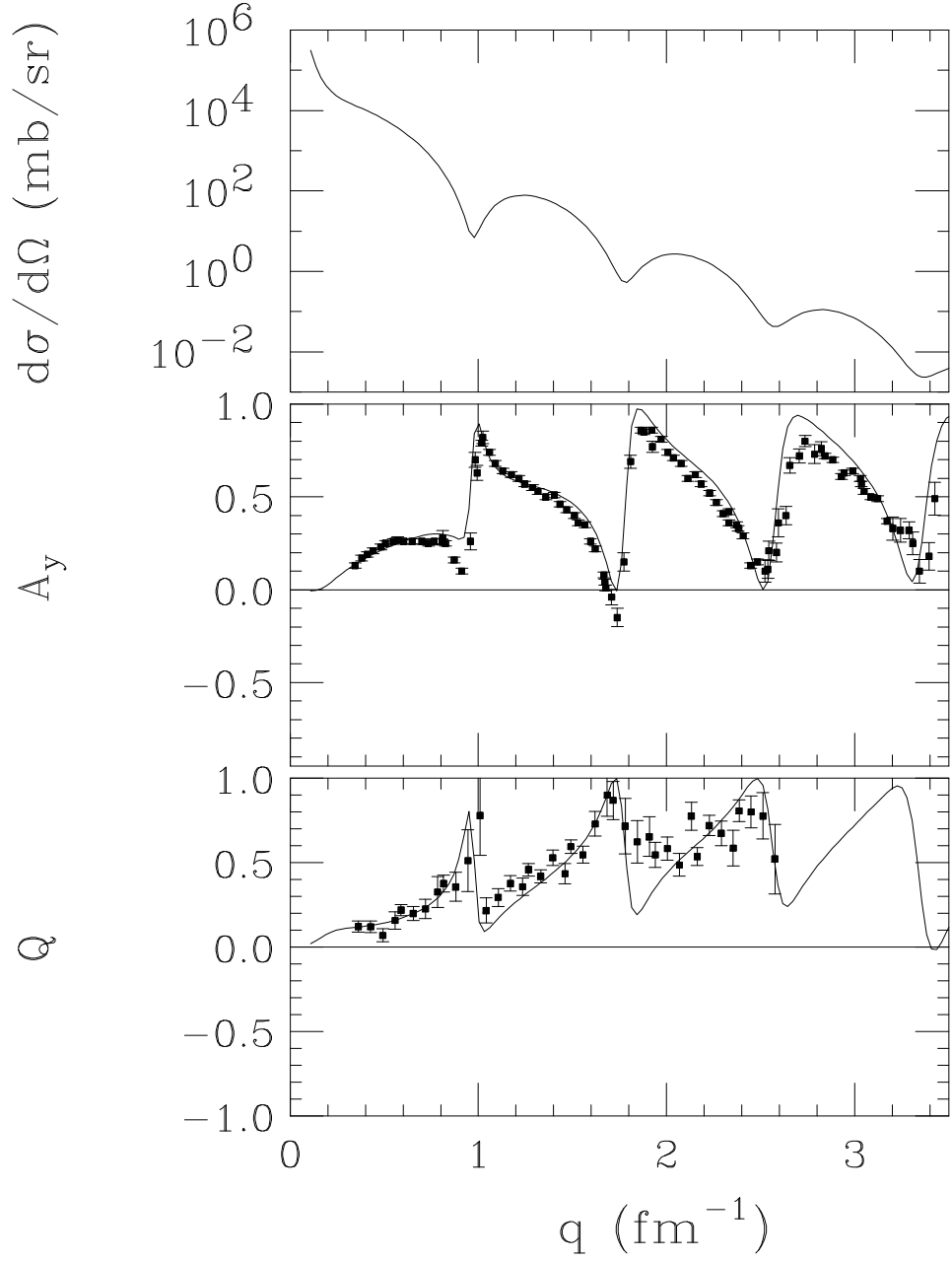}
\caption{Prediction for $^{40}$Ca at 650 MeV.}
\label{fig:ca40spin}
\end{center}
\end{figure}

\begin{figure}[p!]
\begin{center}
\includegraphics[width=5.0in,angle=180,clip=true]{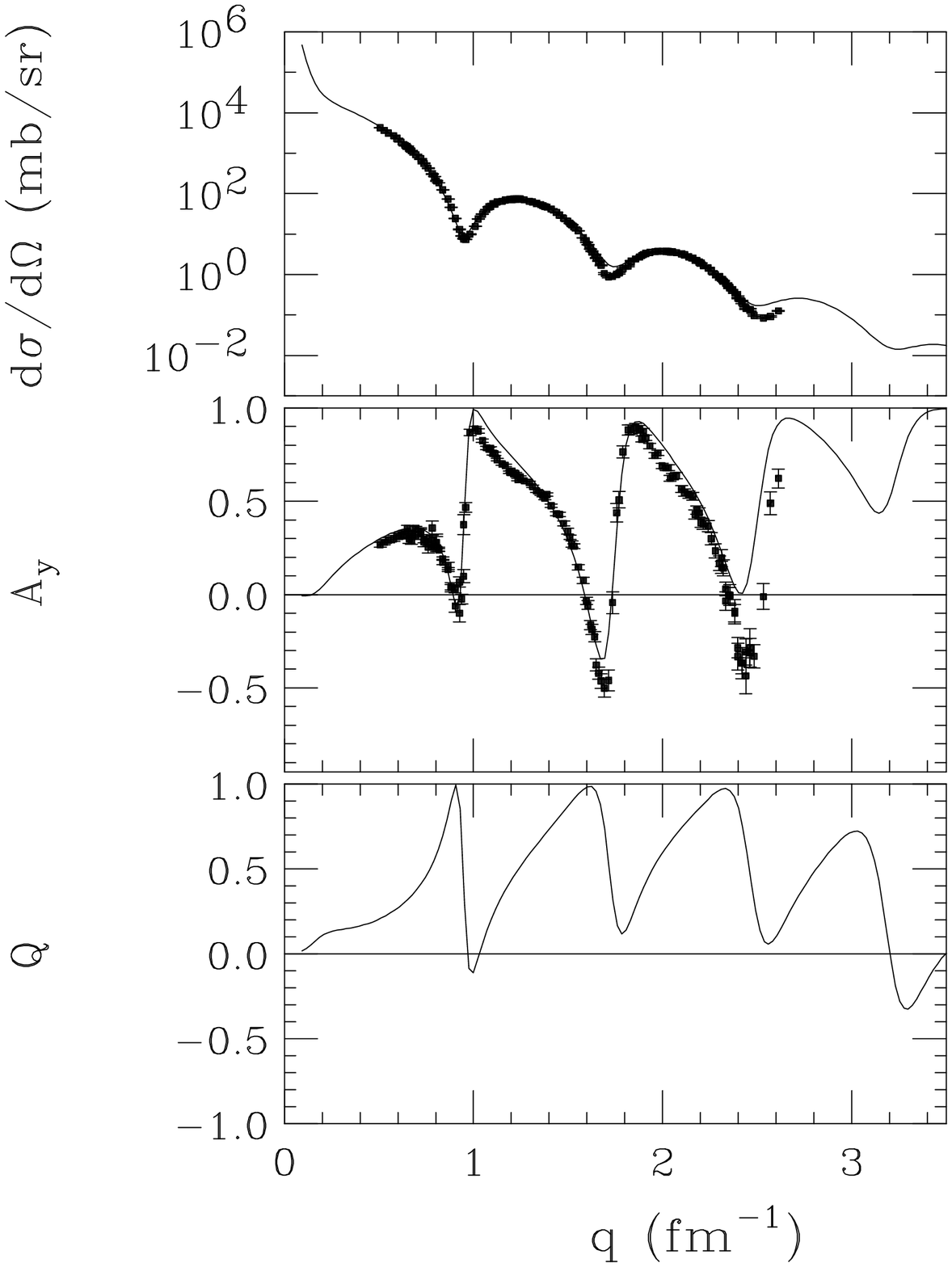}
\caption{Results of the global analysis for $^{48}$Ca at 497.5 MeV.}
\label{fig:ca48spin}
\end{center}
\end{figure}

\begin{figure}[p!]
\begin{center}
\includegraphics[width=5.0in,angle=180,clip=true]{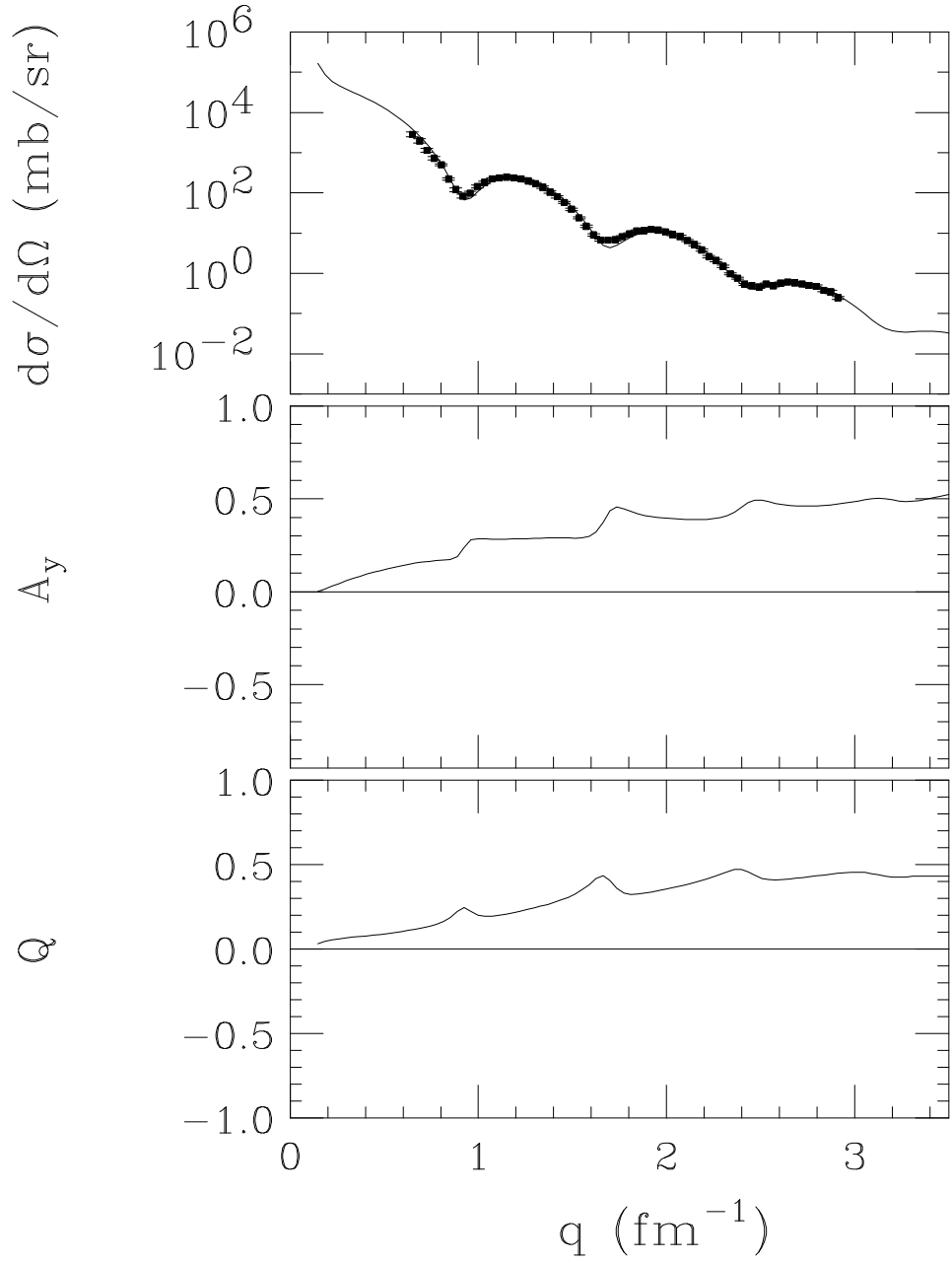}
\caption{Prediction for $^{48}$Ca  at 1044 MeV.}
\label{fig:ca48pred}
\end{center}
\end{figure}

\begin{figure}[p!]
\begin{center}
\includegraphics[width=5.0in,angle=180,clip=true]{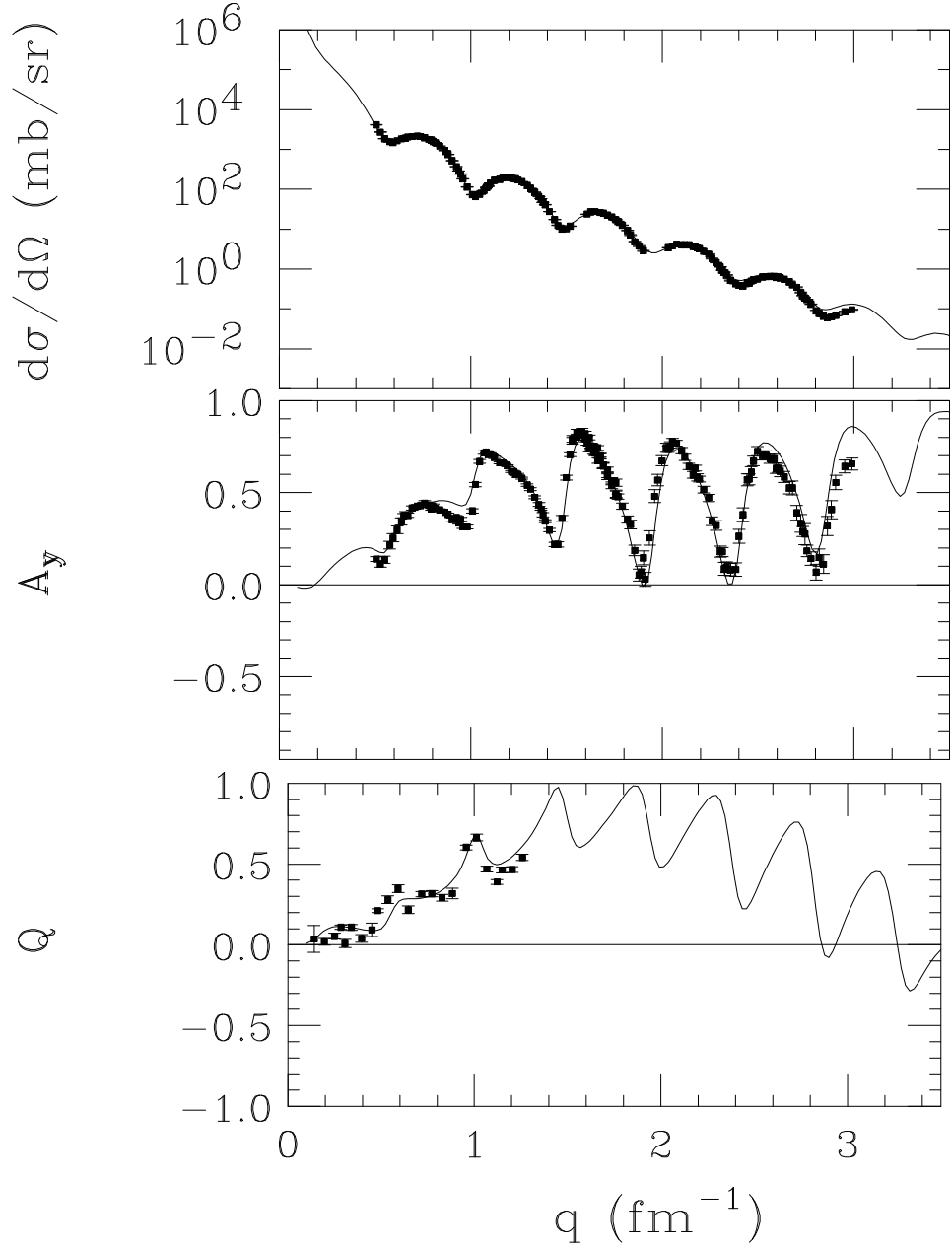}
\caption{Results of the global analysis for $^{208}$Pb at 497.5 MeV.}
\label{fig:pb497spin}
\end{center}
\end{figure}

\begin{figure}[p!]
\begin{center}
\includegraphics[width=5.0in,angle=180,clip=true]{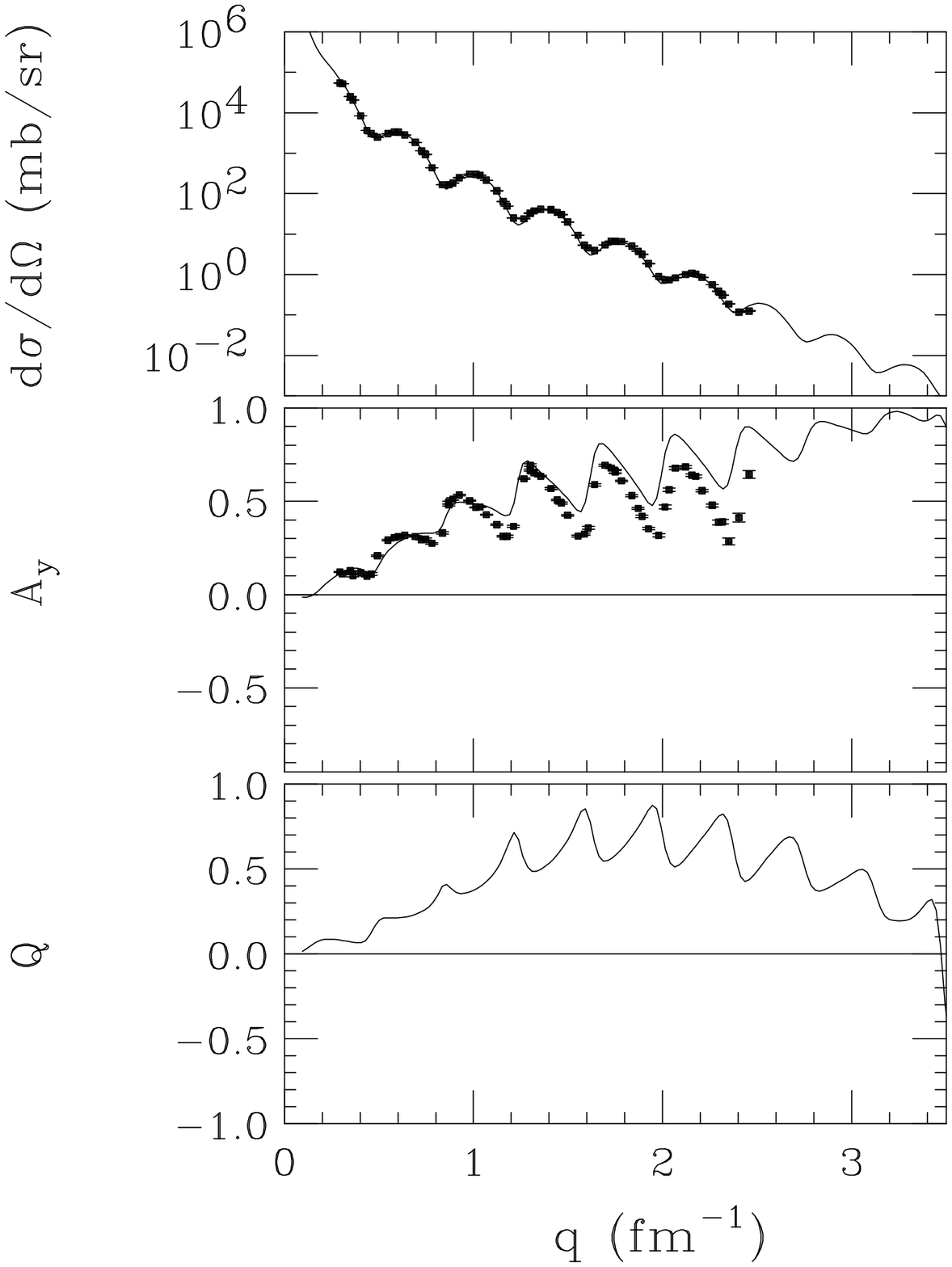}
\caption{Prediction for $^{208}$Pb at 650 MeV}
\label{fig:pb650spin}
\end{center}
\end{figure}

\begin{figure}[p!]
\begin{center}
\includegraphics[width=5.1in,angle=180,clip=true]{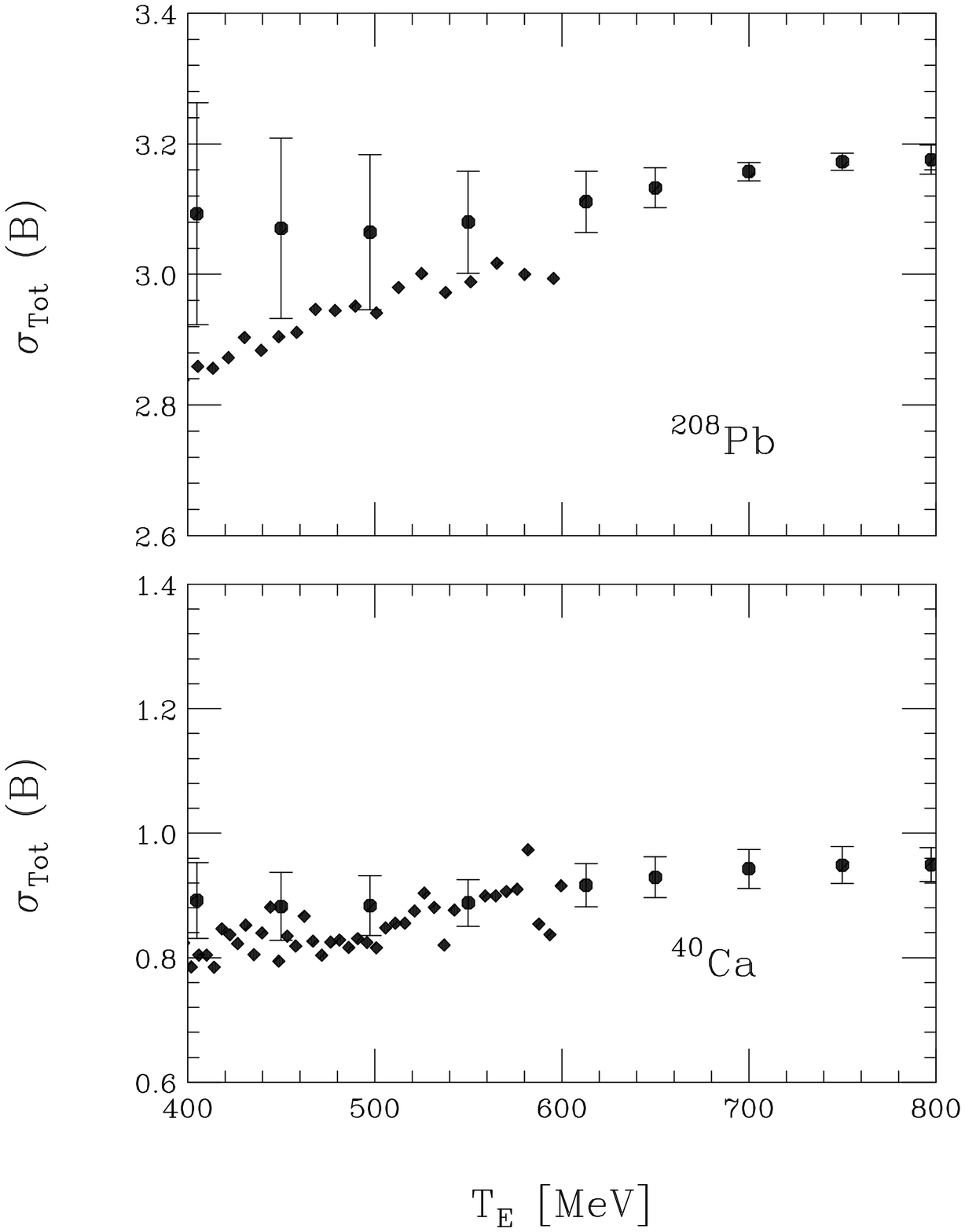}
\caption{ Predicted total cross sections for $^{40}$Ca and $^{208}$Pb
shown as circles are compared with the experimental values 
shown as diamonds from Ref.~\cite{fin}.}

\label{fig:ncapbxs}
\end{center}
\end{figure}

\begin{figure}[p!]
\begin{center}
\includegraphics[width=5.1in,angle=180,clip=true]{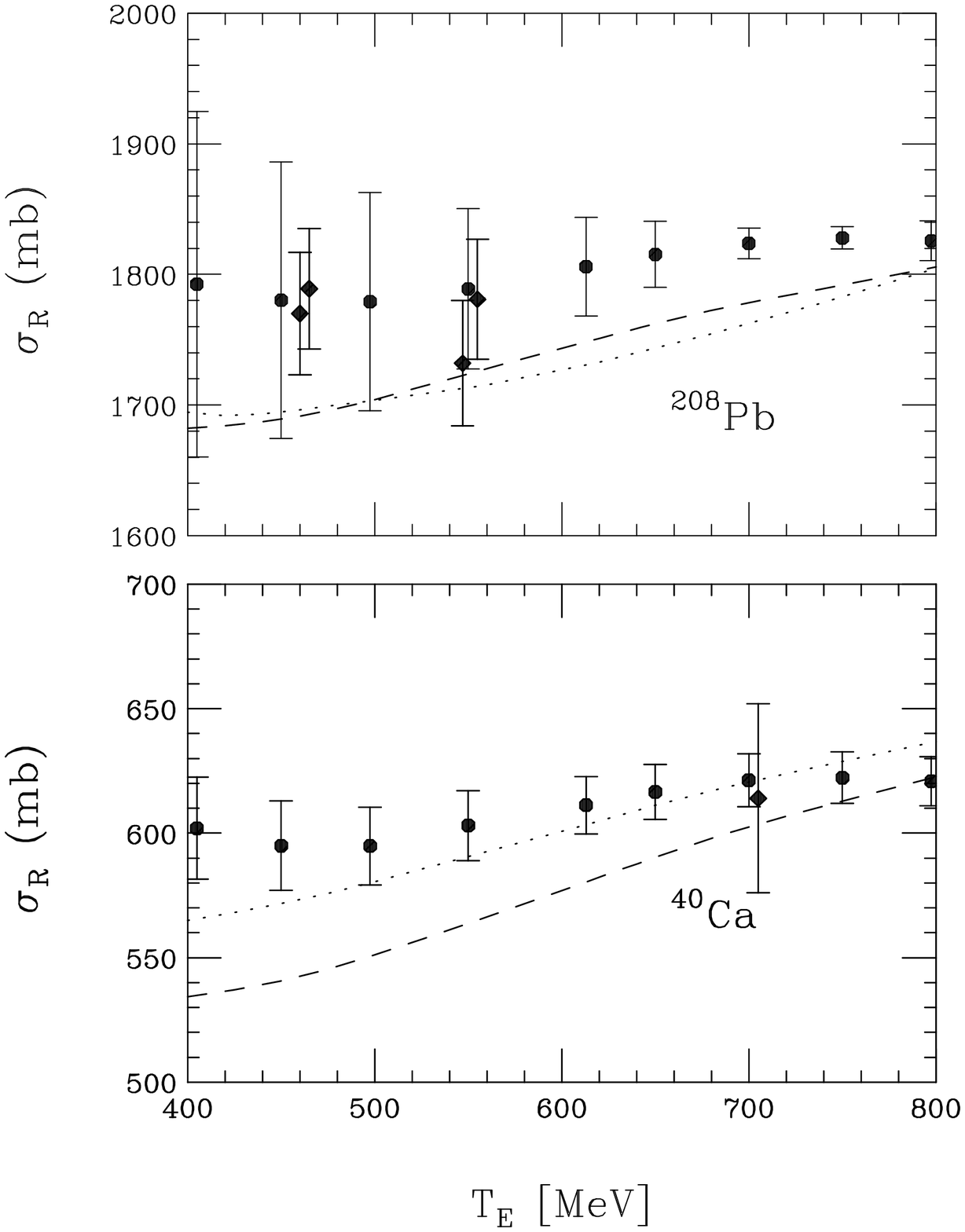}
\caption{ Predicted proton reaction cross sections are shown as  
circles for $^{40}$Ca and $^{208}$Pb 
compared with the experimental values shown as diamonds from Ref.~\cite{rcs}.
The dashed lines are the results of the EDAI global fit and the dotted
lines are the EDAD fit 3 from Ref.~\cite{glob1}.}

\label{fig:reactxs}
\end{center}
\end{figure}

We also predict the total neutron cross section and the 
proton reaction cross section. 
In Fig.~\ref{fig:ncapbxs} the predicted total neutron 
cross sections for $^{40}$Ca and $^{208}$Pb is compared with the 
experimental values from Finlay {\it et al.} \cite{fin}.
The predicted proton reaction cross sections for 
the same two targets are given in Fig.~\ref{fig:reactxs}
with the experimental values \cite{rcs} and the predictions are reasonable.

All of the results confirm that the global approach produces 
good fits to the data, that predictions are quite
acceptable and that the neutron densities are energy independent.  
These densities should be useful input to a large 
number of nuclear reactions. They provide  
the empirical values of $R_n$ and $S_n$  that are needed for testing  
theoretical models.

\section{Summary and Conclusions}
\label{sec:sumcon}

This paper reports a new method for extracting neutron densities from 
intermediate energy elastic proton-nucleus scattering observables. 
Neutron densities are needed for atomic parity violation experiments,
the analysis of antiproton atoms, the experiments parity violation 
elastic electron scattering, and theoretical nonrelativistic and covariant 
mean-field models. It is interesting to note that our results for $S_n$
are in agreement with work using antiprotonic atoms, see Fig.~4 in 
Ref.~\cite{trz}.

The approach uses a global analysis, similar to the
global fits using a Dirac phenomenology, 
but in this case it is based on the relativistic impulse approximation 
(RIA). The input to the procedure are the vector point proton density 
distribution, $\rho^{p}_{v}$, which
is determined from the empirical charge density
after unfolding the proton form factor and the Arndt NN amplitudes.
The other densities, $\rho^{n}_{v}$, $\rho^{p}_{s}$, $\rho^{n}_{s}$,
are parametrized. 
The neutron densities for $^{40}$Ca, $^{48}$Ca and
$^{208}$Pb obtained from the global fits have been tested for sensitivity
to the input.  Different proton form factors, different charge densities,
different sets of the Arndt NN amplitudes, different EFT models, 
different data sets included in the fits, and differences in the
ranges of the momentum transfer  used in the fits have been investigated.
With the exception of the fourier-bessel charge density for $^{40}$Ca
all of these tests produce values for $R_n$ and $S_n$ that
overlap. The prediction of data sets not used in the fits are well reproduced 
and the calculated observables are in good agreement with data.

In conclusion, we have obtained values of $R_n$ and $S_n$ for 
$^{40}$Ca, $^{48}$Ca, and  $^{208}$Pb. 
We have obtain good global fits for MA4, FZ4 and VA3 for
every charge distribution used; SOG, 3PF, and FB. Both five 
momentum transfer ranges and four momentum transfer ranges
were used. As discussed earlier, for each case we calculate 
the mean and standard deviation for $R_n$ and $S_n$ for every global fit, 
this gives us an average over all EFT models for a given charge 
distribution. This is denoted by $AVE_{EFT}$ in the tables. 
As mentioned several times this is not to be taken as a statistical 
analysis, but as a way to show the range of the values in a consistent 
way, they are shown as bars in the figures. 
We do this rather than remove outliers.  
Then for each case we show the combined values of the range for 
$R_n$ and $S_n$ for that case, this gives the average over all 
charge distributions as well as
over all EFT models. These results are denoted as $AVE_{CD}$ and $AVE_{EFT}$ 
in the tables. We then calculate the ranges of $R_n$ and $S_n$
for all three cases combined for each target.
From these final test cases, shown at the bottom of the tables in the
appendix, using the five momentum transfer ranges
we obtain the following values:
for $^{40}$Ca,  $3.350> R_n >3.300$  fm and $-0.008> S_n >-0.080$ fm;
for $^{48}$Ca, $3.505> R_n >3.421$ fm and $0.148> S_n >0.058$ fm;
for $^{208}$Pb, $5.589> R_n >5.513$ and $0.156> S_n >0.076$ fm.
In conclusion, the RMS neutron radius, $R_n$ and the neutron skin
thickness $S_n$ obtained from the global fits
that we consider to be the most reliable ranges of our results
are given as follows: for $^{40}$Ca,  $3.314> R_n >3.310$ fm and
 $-0.063> S_n >-0.067$ fm; for $^{48}$Ca,
$3.459> R_n >3.413$ fm and $0.102> S_n >0.056$ fm; and
for $^{208}$Pb $5.550> R_n >5.522$ fm and $0.111> S_n >0.083$ fm.

The authors have found that the global fits are more stable
when using the Arndt NN amplitude set FA00 rather than the
Arndt NN amplitude set SM86. This might be expected as the FA00 analysis
is more recent. Both of the form factors used, $G_1(q)$ and 
$G_2(q)$ when using FA00 were the most stable,
see Tables~I, II and III in the Appendix.
The impact on the fits due to the charge distribution input, SOG, 3PF and FB, 
as shown in Fig.~\ref{fig:diffcd1}, clearly showed that the cases using the 
SOG and 3PF charge distributions were stable, {\it i.e.} the ranges overlapped,
for all targets, however, for $^{40}$Ca the results for the FB
charge distribution gave very different results, see Table~I.
We also found that the fits using Arndt NN amplitude set FA00,
the $G_1(q)$ form factor, and the SOG charge distribution produced the
most stable results. In this case the values of $R_n$ and $S_n$ using 
the five momentum transfer ranges are:
for $^{40}$Ca, $3.318> R_n >3.308$ fm and $-0.059> S_n >-0.069$ fm;
for $^{48}$Ca, $3.498> R_n >3.412$ fm and $0.141> S_n >0.055$ fm;
and for $^{208}$Pb $5.602> R_n >5.512$ fm and  $0.164> S_n >0.074$ fm.
These values, as well as the values using the most
conservative errors and the results that we consider most  
reliable, are in agreement with nonrelativistic Skyrme models 
\cite{Brown,Brown2,Brown3}, relativistic Hartree-Bogoliubov
model extended to include density dependent meson-nucleon couplings
\cite{Ring2} and results from a recent analysis of 
antiprotonic atoms \cite{trz}. Our results for $^{48}$Ca and $^{208}$Pb 
are generally not in agreement with relativistic mean-field models,
see Ref.~6 and references there in.

The global fit based on the RIA is a new tool for obtaining
the neutron density. The values of $R_n$ and $S_n$ obtained are robust.
This quality is verified in several ways as discussed in Section III.
For example, we checked the sensitivity due to the data sets included in the
fit by doing global fits using only two data sets (497.5 MeV and 797.5 MeV)
for $^{40}$Ca and $^{208}$Pb as well as four data sets.
The values of $S_n$ for the two data set case are:
for $^{40}$Ca $-0.054> S_n >-0.066$ fm; and for $^{208}$Pb 
$0.166> S_n >0.088$ fm. These results agree well with $S_n$ for the 
four data set case: for $^{40}$Ca $-0.059> S_n >-0.069$ fm; and 
for $^{208}$Pb $0.164> S_n >0.074$ fm.
These results, shown in Fig.~\ref{fig:difftwosets}, motivates us to use the
global procedure for all nuclei that have at least two data sets
in the medium energy range.

This work provides energy independent values for $R_n$ and 
$S_n$, in contrast to the energy dependent values obtained by previous studies.
In addition, the results presented in paper show that the expected rms neutron
radius and skin thickness for $^{40}$Ca is accurately reproduced.
The values of $R_n$ and $S_n$ obtained from the global fits
that we consider to be the most reliable are given as follows:  
We plan to  extend our work to additional nuclei and will  
continue to investigate different models and procedures. The goal is to 
continue to improve quality of neutron densities that result from our
global fits.

\vspace*{-2pt}

\acknowledgments

We thank R. J. Furnstahl, C. J. Horowitz and B. D. Serot for useful 
discussions. S. Hama also thanks the Department of Physics, The Ohio State
University for their support while some of this work was done.
B. C. Clark and S. Hama thanks the National Institute for
Nuclear Theory for their hospitality.
This work was supported in part by the National Science Foundation
under Grants No. PHY--9800964 and PHY--0098645.

\hfill

\hfill
\newpage
\appendix*
\setlength{\baselineskip}{1pt}
\begingroup
\squeezetable
\begin{table}
\begin{center}
\caption{}
\begin{tabular}{lcc}  
\hline\\
Range~(fm$^{-1}$)  &     $R_n$~(fm)   &  $S_n$ (fm)\\
& $^{40}Ca$ Case 1 &\\
\hline \\
         &    $G_1(q)$     &      FA00\\
\hline \\
         &       SOG          &     $AVE_{EFT}$\\
2.0-3.5  &  $3.314> R_n >3.310$  & $-0.063> S_n >-0.067$\\
1.5-3.5  &  $3.318> R_n >3.308$  & $-0.059> S_n >-0.069$\\
         &       3PF          &     $AVE_{EFT}$\\
2.0-3.5  &  $3.328> R_n >3.304$  & $-0.051> S_n >-0.075$\\
1.5-3.5  &  $3.328> R_n >3.304$  & $-0.051> S_n >-0.075$\\
         &       FB           &     $AVE_{EFT}$\\
2.0-3.5  &  $3.373> R_n >3.349$  & $0.028> S_n > 0.004$\\
1.5-3.5  &  $3.371> R_n >3.347$  & $0.025> S_n > 0.001$\\
         &     $AVE_{CD}$     &     $AVE_{EFT}$\\
2.0-3.5  &  $3.354> R_n >3.306$  & $0.002> S_n >-0.076$\\ 
1.5-3.5  &  $3.352> R_n >3.306$  & $0.000> S_n >-0.076$\\ 
\hline \\
& $^{40}Ca$ Case 2 &\\
\hline \\
         &    $G_2(q)$     &      FA00\\
\hline \\
         &       SOG          &     $AVE_{EFT}$\\
2.0-3.5  &  $3.314> R_n >3.300$ & $-0.070> S_n >-0.084$\\
1.5-3.5  &  $3.315> R_n >3.299$ & $-0.069> S_n >-0.085$\\
         &       3PF          &     $AVE_{EFT}$\\
2.0-3.5  &  $3.317> R_n >3.299$ & $-0.070> S_n >-0.088$\\
1.5-3.5  &  $3.324> R_n >3.296$ & $-0.062> S_n >-0.090$\\
         &       FB           &     $AVE_{EFT}$\\
2.0-3.5  &  $3.341> R_n >3.333$ & $-0.012> S_n >-0.020$\\
1.5-3.5  &  $3.340> R_n >3.332$ & $-0.013> S_n >-0.021$\\
         &    $AVE_{CD}$      &     $AVE_{EFT}$\\
2.0-3.5  &  $3.333> R_n >3.301$ & $-0.027> S_n >-0.087$\\
1.5-3.5  &  $3.334> R_n >3.302$ & $-0.026> S_n >-0.086$\\
\hline \\
& $^{40}Ca$ Case 3 &\\
\hline\\
         &    $G_1(q)$     &      SM86\\
\hline\\ 
        &       SOG          &     $AVE_{EFT}$\\
2.0-3.5 &  $3.344> R_n >3.308$ & $-0.033> S_n >-0.069$\\
1.5-3.5 &  $3.341> R_n >3.305$ & $-0.036> S_n >-0.072$\\
        &       3PF          &      $AVE_{EFT}$\\
2.0-3.5 &  $3.332> R_n >3.302$ & $-0.047> S_n >-0.077$\\
1.5-3.5 &  $3.330> R_n >3.292$ & $-0.049> S_n >-0.087$\\
        &       FB           &      $AVE_{EFT}$\\
2.0-3.5 &  $3.374> R_n >3.340$ & $0.028> S_n >-0.006$\\
1.5-3.5 &  $3.371> R_n >3.335$ & $0.025> S_n >-0.011$\\
        &    $AVE_{CD}$      &      $AVE_{EFT}$\\
2.0-3.5 &  $3.356> R_n >3.308$ & $0.002> S_n >-0.070$\\
1.5-3.5 &  $3.354> R_n >3.304$ & $-0.001> S_n >-0.075$\\
\hline \\
& $^{40}Ca$ for all three cases &\\
\hline\\
2.0-3.5 &  $3.350> R_n >3.304$ & $-0.006> S_n >-0.080$\\
1.5-3.5 &  $3.350> R_n >3.300$ & $-0.008> S_n >-0.080$\\

\hline
\end{tabular}
\end{center}
\end{table}
\endgroup

\begingroup
\squeezetable
\begin{table}
\begin{center}
\caption{}
\begin{tabular}{lcc}  
\hline\\
Range~(fm$^{-1}$)  &     $R_n$~(fm)   &  $S_n$ (fm)\\
& $^{48}Ca$ Case 1 &\\
\hline \\
         &    $G_1(q)$     &      FA00\\
\hline \\
         &       SOG          &     $AVE_{EFT}$\\
2.0-3.5 &  $3.459> R_n >3.413$ & $0.102> S_n >0.056$\\
1.5-3.5 &  $3.498> R_n >3.412$ & $0.141> S_n >0.055$\\
         &       3PF          &     $AVE_{EFT}$\\
2.0-3.5 &  $3.456> R_n >3.412$ & $0.087> S_n >0.043$\\
1.5-3.5 &  $3.501> R_n >3.409$ & $0.132> S_n >0.040$\\
         &       FB           &     $AVE_{EFT}$\\
2.0-3.5  &  $3.460> R_n >3.420$ & $0.113> S_n >0.073$\\
1.5-3.5  &  $3.499> R_n >3.417$ & $0.152> S_n >0.070$\\
         &     $AVE_{CD}$     &     $AVE_{EFT}$\\
2.0-3.5  &  $3.459> R_n >3.415$ & $0.103> S_n >0.055$\\
1.5-3.5  &  $3.499> R_n >3.413$ & $0.142> S_n >0.054$\\
\hline \\
& $^{48}Ca$ Case 2 &\\
\hline \\
         &    $G_2(q)$     &      FA00\\
\hline \\
         &       SOG          &     $AVE_{EFT}$\\
2.0-3.5  &  $3.453> R_n >3.407$ & $0.089> S_n >0.043$\\
1.5-3.5  &  $3.498> R_n >3.404$ & $0.134> S_n >0.040$\\
         &       3PF          &     $AVE_{EFT}$\\
2.0-3.5  &  $3.448> R_n >3.404$ & $0.072> S_n >0.028$\\
1.5-3.5  &  $3.500> R_n >3.398$ & $0.124> S_n >0.022$\\
         &       FB           &     $AVE_{EFT}$\\
2.0-3.5 &   $3.458> R_n >3.414$ & $0.103> S_n >0.059$\\
1.5-3.5 &   $3.511> R_n >3.407$ & $0.157> S_n >0.053$\\
         &    $AVE_{CD}$      &     $AVE_{EFT}$\\
2.0-3.5 &   $3.454> R_n >3.408$ & $0.092> S_n >0.040$\\
1.5-3.5 &   $3.503> R_n >3.403$ & $0.139> S_n >0.037$\\
\hline \\
& $^{48}Ca$ Case 3 &\\
\hline\\
         &    $G_1(q)$     &      SM86\\
\hline\\ 
        &       SOG          &     $AVE_{EFT}$\\
2.0-3.5  &  $3.506> R_n >3.476$ & $0.150> S_n >0.120$\\
1.5-3.5  &  $3.505> R_n >3.463$ & $0.148> S_n >0.106$\\
        &       3PF          &      $AVE_{EFT}$\\
2.0-3.5  &  $3.487> R_n >3.447$ & $0.119> S_n >0.079$\\
1.5-3.5  &  $3.484> R_n >3.446$ & $0.116> S_n >0.078$\\
        &       FB           &      $AVE_{EFT}$\\
2.0-3.5  &  $3.516> R_n >3.486$ & $0.169> S_n >0.139$\\
1.5-3.5  &  $3.515> R_n >3.473$ & $0.167> S_n >0.125$\\
        &    $AVE_{CD}$      &      $AVE_{EFT}$\\
2.0-3.5  &  $3.509> R_n >3.465$ & $0.158> S_n >0.100$\\
1.5-3.5  &  $3.504> R_n >3.458$ & $0.152> S_n >0.094$\\
\hline \\
& $^{48}Ca$ for all three cases &\\
\hline\\
2.0-3.5 &  $3.485> R_n >3.417$ &  $0.129> S_n >0.053$\\
1.5-3.5 &  $3.505> R_n >3.421$ &  $0.148> S_n >0.058$\\

\hline
\end{tabular}
\end{center}
\end{table}
\endgroup

\begingroup
\squeezetable
\begin{table}
\begin{center}
\caption{}
\begin{tabular}{lcc}  
\hline\\
Range~(fm$^{-1}$)  &     $R_n$~(fm)   &  $S_n$ (fm)\\
& $^{208}Pb$ Case 1 &\\
\hline \\
         &    $G_1(q)$     &      FA00\\
\hline \\
         &       SOG          &     $AVE_{EFT}$\\
2.0-3.5 &  $5.550> R_n >5.522$ &  $0.111> S_n >0.083$\\
1.5-3.5 &  $5.602> R_n >5.512$ &  $0.164> S_n >0.074$\\
         &       3PF          &     $AVE_{EFT}$\\
2.0-3.5 &  $5.580> R_n >5.548$ &  $0.155> S_n >0.123$\\
1.5-3.5 &  $5.586> R_n >5.552$ &  $0.160> S_n >0.126$\\
         &       FB           &     $AVE_{EFT}$\\
2.0-3.5 &  $5.546> R_n >5.518$ &  $0.108> S_n >0.080$\\
1.5-3.5 &  $5.600> R_n >5.508$ &  $0.162> S_n >0.070$\\
         &     $AVE_{CD}$     &     $AVE_{EFT}$\\
2.0-3.5 &  $5.565> R_n >5.523$ &  $0.135> S_n >0.085$\\
1.5-3.5 &  $5.599> R_n >5.521$ &  $0.167> S_n >0.085$\\
\hline \\
& $^{208}Pb$ Case 2 &\\ 
\hline \\
         &    $G_2(q)$     &      FA00\\
\hline \\
         &       SOG          &     $AVE_{EFT}$\\
2.0-3.5 &  $5.545> R_n >5.517$  &  $0.102> S_n >0.074$\\  
1.5-3.5 &  $5.599> R_n >5.507$  &  $0.156> S_n >0.064$\\  
         &       3PF          &     $AVE_{EFT}$\\
2.0-3.5 &  $5.576> R_n >5.538$   & $0.146> S_n >0.108$\\
1.5-3.5 &  $5.582> R_n >5.542$  &  $0.152> S_n >0.112$\\
         &       FB           &     $AVE_{EFT}$\\
2.0-3.5 &  $5.537> R_n >5.511$  &  $0.095> S_n >0.069$\\
1.5-3.5 &  $5.593> R_n >5.499$  &  $0.151> S_n >0.057$\\
         &    $AVE_{CD}$      &     $AVE_{EFT}$\\
2.0-3.5 &  $5.558> R_n >5.516$  &  $0.124> S_n >0.074$\\
1.5-3.5 &  $5.594> R_n >5.514$  &  $0.157> S_n >0.075$\\
\hline \\
& $^{208}Pb$ Case 3 &\\ 
\hline\\
         &    $G_1(q)$     &      SM86\\
\hline\\ 
        &       SOG          &     $AVE_{EFT}$\\
2.0-3.5 &  $5.538> R_n >5.498$  &  $0.100> S_n >0.060$\\
1.5-3.5 &  $5.554> R_n >5.502$  &  $0.115> S_n >0.063$\\
        &       3PF          &      $AVE_{EFT}$\\
2.0-3.5 &  $5.584> R_n >5.532$  &  $0.158> S_n >0.106$\\
1.5-3.5 &  $5.590> R_n >5.538$  &  $0.165> S_n >0.113$\\
        &       FB           &      $AVE_{EFT}$\\
2.0-3.5 &  $5.532> R_n >5.502$  &  $0.094> S_n >0.064$\\
1.5-3.5 &  $5.551> R_n >5.503$  &  $0.113> S_n >0.065$\\
        &    $AVE_{CD}$      &      $AVE_{EFT}$\\
2.0-3.5 &  $5.559> R_n >5.503$  &  $0.129> S_n >0.065$\\
1.5-3.5 &  $5.571> R_n >5.509$  &  $0.141> S_n >0.071$\\
\hline \\
& $^{208}Pb$ for all three cases &\\

\hline\\
2.0-3.5 &  $5.561> R_n >5.513$  &  $0.130> S_n >0.074$\\
1.5-3.5 &  $5.589> R_n >5.513$ &   $0.156> S_n >0.076$\\

\hline
\end{tabular}
\end{center}
\end{table}
\endgroup

\end{document}